\documentclass[10pt]{article}

\newcommand{\blind}{1}

\usepackage{listings}
\usepackage{cancel}
\usepackage[normalem]{ulem}
\usepackage{comment}
\usepackage{float}
\usepackage{array,multirow}
\usepackage{placeins}
\usepackage{epsfig}
\usepackage[english]{babel}
\usepackage{graphicx}
\usepackage{setspace}
\usepackage{amsmath}
\usepackage{relsize}
\usepackage{color,soul}
\usepackage{natbib}
\usepackage{makeidx}
\usepackage[caption = false]{subfig}
\usepackage{multicol}
\usepackage{setspace}

\begin{document}
	
	\def\spacingset#1{\renewcommand{\baselinestretch}%
		{#1}\small\normalsize} \spacingset{1}
	
	\if1\blind
	{
		\title{\bf Decision Making in\\Drug Development via\\ {Inference on Power}\\
}
		\author{Geoffrey S Johnson \vspace{2mm}\\
    			Merck \& Co., Inc. \\
			   770 Sumneytown Pike, West Point, PA 19438 USA\\
			   geoffrey.s.johnson@gmail.com\vspace{2mm}
	      }
		\date{}
		\maketitle
	} \fi
	
	\if0\blind
	{
		\title{\bf Decision Making in\\ Drug Development via\\ Inference on Power}
		\date{}		
		\maketitle
	} \fi

	\title{}

	\label{firstpage}

	\begin{abstract}

A typical power calculation is performed by replacing unknown population-level quantities in the power function with what is observed in external studies.  Many authors and practitioners view this as an assumed value of power and offer the Bayesian quantity probability of success or assurance as an alternative.  The claim is by averaging over a prior or posterior distribution, probability of success transcends power by capturing the uncertainty around the unknown true treatment effect and any other population-level parameters.    We use {p-value functions} to frame both the probability of success calculation and the typical power calculation as merely producing two different point estimates of power.  We demonstrate that Go/No-Go decisions based on either point estimate of power do not adequately quantify and control the risk involved, and instead we argue for Go/No-Go decisions that utilize inference on power for better risk management and decision making.  

	\end{abstract}

	\noindent
	{\it Keywords:}  Pharmaceutical drug development, P-value function, Confidence distribution,  Probability of success, Assurance.
	\vfill
	
	\pagebreak
	\section{Introduction}	
The need for quantitative decision rules in the pharmaceutical industry across all phases of clinical development is paramount (\citeauthor{frewer2016}  \citeyear{frewer2016}; \citeauthor{kirby2017} \citeyear{kirby2017}; \citeauthor{lalonde2007} \citeyear{lalonde2007}).  
 This entails Go/No-Go decisions from phase 1 through 3, and just as important is the probability of making these decisions.  In drug development many authors propose Bayesian predictive probability as a more appropriate alternative to frequentist power, be it for interim analyses or across phases of development, and espouse its use as part of net present value calculations (\citeauthor{o2005} 2005; \citeauthor{trzaskoma2007} 2007; \citeauthor{chuang2006} 2006).  The claim is that one must assume a particular parameter value (population-level treatment effect) is true in order to calculate power, whereas a Bayesian approach considers the parameter itself as a random variable, so that Bayesian \textit{probability of success} exists unconditionally on the parameter of interest (\citeauthor{temple2021} 2021; \citeauthor{best2018} 2018; \citeauthor{ciarleglio2017} 2017; \citeauthor{king2009} 2009).  Examples abound comparing probability of success calculations to misguided evaluations of the power curve as evidence that power is overly optimistic or anti-conservative when used in decision making (\citeauthor{saville2014} 2014).  While there is certainly value in predicting a clinical trial result, 
the frequentist confidence or Bayesian credible level associated with a prediction interval 
is not a 
statement solely about the 
clinical trial itself.  Viewed this way,  the Bayesian term 
\textit{probability of success} is a misnomer and may not be the primary quantity of interest for decision making in drug development.  This confusion is due in large part to the relaxed definition of probability used in Bayesian inference where a parameter (e.g. the population-level treatment effect) is treated as an unrealized or unobservable realization of a random variable that depends on  observed data, and probability is reinterpreted as measuring the freely assigned unfalsifiable belief of the experimenter.   
The key to appreciate our approach for decision making is adopting an objective definition of probability $-$ although we do not know the population-level parameter of interest, this does not mean it is a random variable, and our estimation, inference, and decision making should not treat it as random.  A major focus of this manuscript is to frame power not as an assumed parameter, but as a parameter that one can estimate and infer, and to demonstrate that Bayesian \textit{probability of success} is not a ``fix'' for power.  An excellent critique of probability of success has been provided by \citeauthor{carroll2013} (2013) who offers a summary of its features using a simple normal model, and an example involving a hazard ratio while considering that the phase 2 posterior is centered at the unknown fixed true treatment effect to be investigated in phase 3.  This has incredible value for understanding the properties of probability of success, but their investigation inherently treats probability of success as a population-level quantity that exists in addition to power.  Our contribution is to build on this discussion by interpreting probability of success as a point estimate of power, and to argue in favor of Go/No-Go decisions that instead utilize a transformation-invariant estimate of power as well as inference on power.  The most critical point we demonstrate is that if inference on power is ignored, the decision maker may otherwise be indifferent and unwittingly exposed to risk when choosing programs to progress to phase 3 based on point estimates of power.  
\\

Bayesian statements of belief probability are visually depicted through prior and posterior distributions, \textit{distribution estimates} of an unknown quantity of interest, and are powerful tools for visualizing and pooling prior information and expert opinion with current data.  
\cite{spiegelhalter2004} illustrate this and highlight its application to forming stopping rules for early efficacy, futility, and safety, as well as planning future studies.  Under the frequentist paradigm, the analogous distribution estimate is a \textit{confidence curve} composed of 
{\textit{p-value functions}}, a sample-dependent ex-post object that depicts all possible p-values and confidence intervals one could construct for a parameter of interest, given the observed data.  
  This {confidence curve} has the appearance of a folded cumulative distribution function over the parameter space, but it does not depict a random parameter.  Instead, it summarizes all possible inference one could perform based on a given data set using a particular hypothesis test or confidence interval method.  {P-value functions} and confidence curves allow for meta-analysis (\citeauthor{xie2011confidence} 2011) and can even be used to capture and incorporate expert opinion by considering a hypothetical experimental result (\citeauthor{xie2013incorporating} 2013), providing a powerful visual tool for decision making across all phases of clinical development.  {When the p-value is uniformly distributed under the null and the p-value function has the appearance of a distribution function on the parameter space, it is often referred to as a confidence distribution} (\citeauthor{xie2013} 2013; \citeauthor{schweder2016} 2016).    
\\

The original idea for the confidence distribution dates back to Sir Ronald Fisher, who initially termed it the fiducial or ``faith" distribution.  He viewed the p-value as a continuous measure of evidence, drawing inspiration from Jeffreys' work in objective Bayesianism, and he opposed the Neyman-Wald approach to hypothesis testing.  He also opposed the other end of statistical inference using personal or subjective probabilities championed by Savage and de Finetti (\citeauthor{efron1998} 1998).  Fisher developed likelihood-based inference aiming to combine information from different sources with an emphasis on model coherence and optimality, and intended the fiducial distribution to be a universal approach for Bayesian-like inference in the absence of a prior distribution.  Textbooks and institutions ultimately adopted the Neyman-Wald approach to hypothesis testing, obscuring the true merit of the p-value.  However, in the decades since there has been renewed interest in the topic using a purely frequentist interpretation (\citeauthor{efron1998} 1998), and the confidence distribution has become a remarkable achievement inspired by Fisher.
\\

The novelty of this manuscript is on the interpretation and visualization of statistical inference, the mathematical considerations for 
constructing a {p-value function} for power, and the statistical evaluation of performing inference on power in comparison to existing methods for decision making in drug development.  Section \ref{methods} formally defines a {p-value function} linking it to hypothesis testing and meta-analysis, and extends these developments to inference on power.  
Section \ref{QDM} demonstrates the use of p-value functions in the decision making framework across phases 2 and 3 of pharmaceutical development.  Desired inference on phase 3 power is used to reverse engineer the hypothesis, significance level, and sample size required in phase 2.  In Section \ref{sim section} this approach is evaluated through simulation alongside decision rules using probability of success and a typical power calculation, 
and a discussion is provided in Appendix \ref{multiple comparisons} on why adjustment for multiple comparisons is not required if one adopts a Fisherian point of view.  SAS code is provided in Appendix \ref{sas code}.

\section{Methods}\label{methods}
\subsection{P-value Functions}\label{CDs}

A confidence interval for a parameter $\theta$ is a set of {plausible hypotheses} for $\theta$, given the data $\boldsymbol{X}=\boldsymbol{x}$ observed.  Two well-known and often related methods for producing confidence intervals are inverting a family of hypothesis tests and using a pivotal quantity.  The most familiar example of inverting a hypothesis test utilizes the likelihood ratio.  This test uses the entire full likelihood as the test statistic and answers the question, ``If $\theta$ really is equal to $\theta_0$, how often would we observe a likelihood situated as far away \textit{and} spread out as what we have observed (as measured by the likelihood ratio) or something even more extreme?" Consider initially a simple normal setting where $X_1,...,X_n \sim N(\theta, \sigma)$. Here the likelihood ratio statistic for $\theta$ is a monotonic transformation of the squared t-statistic.  Were it not that the t-distribution is free of the nuisance parameter $\sigma$, this parameter would be replaced with a restricted estimate calculated under the null when using the t-distribution to characterize the performance of the t-statistic and ultimately produce the p-value. Alternatively, {when mild regularity conditions are met} under $H_0$: $\theta=\theta_0$, one could use the $\chi^2_1$ approximation for the sampling distribution of the likelihood ratio test statistic, an approximation free of any nuisance parameters (\citeauthor{wilks1938} 1938).  If an upper-tailed test is applied to all values of $\theta$ in the parameter space, the resulting {function of upper-tailed p-values is called an upper p-value function.}  
The one-sided p-value testing $H_0$: $\theta\le\theta_0$,
\begin{eqnarray}\label{eq}
H(\theta_0,\boldsymbol{x})= \left\{ \begin{array}{cc}
\big[1-F_{\chi^2_1}\big(-2\text{log}LR(\boldsymbol{x},\theta_0)\big)\big]/2 & \text{if } \theta_0 \le \hat{\theta}_{mle} \\
 &  \\
 \big[1+F_{\chi^2_1}\big(-2\text{log}LR(\boldsymbol{x},\theta_0)\big)\big]/2  & \text{if } \theta_0 > \hat{\theta}_{mle}, \end{array}  \right.
\end{eqnarray} 
 as a function of $\theta_0$ and the observed data $\boldsymbol{x}$ is the corresponding {upper p-value function}, where $\hat{\theta}_{mle}$ is the maximum likelihood estimate of $\theta$, $-2\text{log}LR(\boldsymbol{X},\theta_0)$ is the likelihood ratio statistic, and $F_{\chi^2_1}(\cdot)$ is the cumulative distribution function of a $\chi^2_1$ random variable.  Typically the naught subscript is dropped and $\boldsymbol{x}$ is suppressed to emphasize that $H(\theta)$ is a function over the entire parameter space.  Each value in the parameter space takes its turn playing the role of null hypothesis, and hypothesis testing (akin to proof by contradiction) is used to infer the unknown fixed true $\theta$.  This recipe of viewing the p-value as a function of $\theta$ given the data produces a p-value function for any hypothesis test.  For instance, when the sampling distribution of an estimator $g\{\hat{\theta}(\boldsymbol{X})\}$, for some link function $g\{\cdot\}$, is well approximated by a normal distribution, $g\{\hat{\theta}(\boldsymbol{X})\}\sim N\big(g\{\theta\},\hat{\text{se}}\big)$, an upper p-value function for testing hypotheses about $\theta$ is easily produced using a Wald test, $H(\theta)=1-F\big(g\{\hat{\theta}(\boldsymbol{x})\}; g\{\theta\},\hat{\text{se}}\big)=1-\Phi\big([g\{\hat{\theta}(\boldsymbol{x})\}-g\{\theta\}]/\hat{\text{se}}\big)$, where $\hat{\text{se}}$ is a model-based or sandwich estimate for the standard error of $g\{\hat{\theta}(\boldsymbol{X})\}$, $F\big(\cdot; g\{\theta\},\hat{\text{se}}\big)$ is the cumulative distribution function of a normal distribution with mean $g\{\theta\}$ and standard deviation $\hat{\text{se}}$, and $\Phi(\cdot)$ is the cumulative distribution function of the standard normal distribution.  See Appendix \ref{link functions} for further discussion on link functions.  {Without necessarily appealing to regularity conditions and standard asymptotics, one can derive or approximate the sampling distribution of an estimator $g\{\hat{\theta}(\boldsymbol{X})\}$ and utilize its cumulative distribution function (parameterized in terms of its standard error, replaced with a data-driven estimate) to construct an upper p-value function $H(\theta)$ $-$ what could be dubbed the paramter-index CDF method.  This estimation of the standard error can be unrestricted to produce a Wald-type test, or performed under the restricted null space to produce a score-type test.
\\

{The lower p-value function} $H^{-}(\theta)$ can be analogously defined that contains all lower-tailed p-values as a function of $\theta$.  One can then define the confidence curve of one-sided p-values as 
\begin{eqnarray}
C(\theta)= \left\{ \begin{array}{cc}
H(\theta) & \text{if } \theta \le \hat{\theta}(\boldsymbol{x}) \\
 &  \nonumber\\
 H^{-}(\theta)  & \text{if } \theta \ge \hat{\theta}(\boldsymbol{x}). \end{array}  \right.\nonumber
\end{eqnarray}
This definition differs slightly from others (\citeauthor{thornton2020} 2020; \citeauthor{xie2013} 2013; \citeauthor{birnbaum1961} 1961) 
 {and may take on two values at $\theta=\hat{\theta}(\boldsymbol{x})$ forming a jump discontinuity.}  Alternatively, it could be defined as $C(\theta)=\text{min}\{H(\theta), H^{-}(\theta) \}$. The confidence curve can accommodate a discrete sampling distribution where $H(\theta)\ne1-H^{-}(\theta)$, {and it can also accommodate a discrete parameter space.}  The p-value depicts the ex-post sampling probability of the observed result or something more extreme if the hypothesis is true and coincides with the smallest significance level at which the hypothesis would be tentatively ruled out based on the data.  The two-sided equal-tailed $100(1-\alpha)\%$ confidence interval is the set of all hypotheses for $\theta$ such that the smallest one-sided p-value is greater than $\alpha/2$ $-$ i.e., those hypotheses for which the observed result is within a two-sided $100(1-\alpha)\%$ margin of error.  Based on an understanding of this performance through expected gain or expected loss, one would be willing to bet $100(1-\alpha)$ cents to the dollar that the  $100(1-\alpha)\%$ confidence interval computed using observed data indeed covers the unknown fixed true $\theta$.  
\\

{Many times, though not always, the upper p-value function has the appearance of a cumulative distribution function on the parameter space.  In these settings, if the sample space is continuous so that $H(\theta)=1-H^{-}(\theta)$ and the p-value is uniformly distributed under the null, $H(\theta)$ is often referred to as a confidence distribution function and can be depicted by its density $h(\theta)=dH(\theta)/d\theta$} (\citeauthor{xie2013} 2013).  \cite{singh2007} and others highlight an interesting coincidence that when a plug-in estimated sampling distribution or a bootstrap estimated sampling distribution approaches a normal distribution (symmetric shift model) with increasing sample size, it is a valid asymptotic confidence distribution.  Similarly, when a normalized likelihood  (proper Bayesian posterior from improper ``$d\theta$" prior) (\citeauthor{efron1986} 1986) approaches a normal distribution with increasing sample size it too is a valid asymptotic confidence distribution (\citeauthor{fraser2011} 2011; \citeauthor{efron1986} 1986; \citeauthor{xie2013incorporating} 2013; \citeauthor{xie2013} 2013).  {In settings where regular asymptotics do not apply these distribution estimates often still work well as approximate p-value functions.   Even when the p-value is not uniformly distributed under the null or $H(\theta)$ does not necessarily form a distribution function on the parameter space (or both), the p-value function and confidence curve might still be informally called a distribution estimate or a confidence distribution.}  Appendix \ref{definitions} provides the formal definition of a confidence interval (\citeauthor{casella2002} 2002) and confidence distribution function (\citeauthor{xie2013} 2013; \citeauthor{xie2013incorporating} 2013), and an example is discussed in Appendix \ref{discrete example} involving a discrete parameter space.
\\

The p-value function, confidence curve, and confidence density are useful for graphically representing frequentist inference.  They are also useful for performing a meta-analysis.  For pooling prior information with current data, {the p-value from a fixed effect meta-analysis combining two studies} may take the form
\begin{eqnarray}\label{pvalue meta}
p^{(c)}=\Phi\Bigg(\frac{ \frac{1}{\hat{\text{se}}_1}\Phi^{-1}\big(p_1\big) +  \frac{1}{\hat{\text{se}}_2}\Phi^{-1}\big(p_2\big)}{\Big(\frac{1}{\hat{\text{se}}_1^2}+\frac{1}{\hat{\text{se}}_2^2}\Big)^{1/2}}\Bigg),
\end{eqnarray}
where p-values $p_1$ and $p_2$ are back-transformed into $z$-scores, inversely weighted by their corresponding estimated standard errors $\hat{\text{se}}_1$ and $\hat{\text{se}}_2$, and transformed once again into a combined p-value.  $\Phi(\cdot)$ is the cumulative distribution function of the standard normal distribution, and $\Phi^{-1}(\cdot)$ is the corresponding quantile function.
Viewing each p-value as a function of the hypothesis for $\theta$ being tested, this same convolution formula can be applied to p-value functions, i.e.

\begin{eqnarray}\label{conf_dist_2}
H^{(c)}(\theta)=\Phi\Bigg(\frac{ \frac{1}{\hat{\text{se}}_1}\Phi^{-1}\big(H_1(\theta)\big) +  \frac{1}{\hat{\text{se}}_2}\Phi^{-1}\big(H_2(\theta)\big)}{\Big(\frac{1}{\hat{\text{se}}_1^2}+\frac{1}{\hat{\text{se}}_2^2}\Big)^{1/2}}\Bigg).
\end{eqnarray}  
Even in non-normal settings this formula works well to preserve Fisher information (\citeauthor{xie2013incorporating} 2013).  Alternatively, using likelihood-based methods one could multiply the historical and current likelihoods together to form a joint likelihood and use this to conduct a hypothesis test.  
This multiplication of independent likelihoods is precisely what Bayes' theorem accomplishes (plus normalization), without the inversion of a hypothesis test.  In more complicated situations involving a multi-dimensional parameter space, Equations (\ref{pvalue meta}) and (\ref{conf_dist_2}) highlight the notion of division of labor allowing one to avoid construction of an all-encompassing model (\citeauthor{efron1986} 1986).  
\\

{The meta-analytic p-value function above treats the two experiments as a single larger experiment.  When investigating the plausibility of $H_0$: $\theta \le \theta_0$ the meta-analytic p-value could instead be defined as $H^{(c)}(\theta)=H_1(\theta)\cdot H_2(\theta)$.  This treats each experimental result as a separate observation and depicts the upper-tailed probability of observing a result as or more extreme than that witnessed in experiment 1 \textit{and} experiment 2, given hypotheses of the form $H_0$: $\theta \le \theta_0$.  The meta-analytic p-value function of lower-tailed ``\textit{or}'' probability statements testing hypotheses of the form $H_0$: $\theta \ge \theta_0$ can be analogously constructed as $H^{-(c)}(\theta)=H^{-}_1(\theta)+ H^{-}_2(\theta) - H^{-}_1(\theta)\cdot H^{-}_2(\theta)$. } Appendix \ref{construction} provides additional examples showing the construction of a confidence density and its usefulness in a meta-analysis.  It also provides further discussion on Bayesian and frequentist interpretations of probability (\citeauthor{good1965}\nocite{good1966} 1965, 1966; \citeauthor{schrodinger1980} 1980; \citeauthor{ballentine1970} 1970).

\subsection{{Power and Probability of Success}}\label{power and pos}
{The power curve depicts the ex-ante sampling probability of the test statistic (testing a single research hypothesis) as a function of all unknown true population-level parameters.  An understanding of this long-run sampling probability imbues one with a willingness to bet regarding the next experimental result, were the true population-level parameters known.  The power curve is typically constructed while estimating the unknown true population-level nuisance parameters based on a literature review of external studies. (The estimated power curve described above for an upper-tailed test can be approximated using an upper p-value function,.  
See Appendix \ref{approximating power} for further discussion and the SAS code in Appendix \ref{sas code}.)   
\\

A {p-value function} $H(\theta)$ containing inference on $\theta$ from an external study can be used to obtain a {p-value function} for hypotheses concerning the power of a future study.  Since the estimated power function is a monotonic transformation of theta, $\beta(\theta)$, a change of variables in $H(\theta)$ produces a p-value function in terms of power, 
\begin{eqnarray}
H(\theta)=H\big(\beta^{-1}\{\beta(\theta)\}\big),\label{power cd}
\end{eqnarray}
 where $\beta^{-1}$ is the inverse power function.  In practice this can be solved numerically so that the inverse power function is not required.  That is, for a given hypothesis for $\theta$ the value $H(\theta)$ is the p-value function corresponding to $\beta(\theta)$.   This applies the estimated power function to confidence limits for $\theta$ to construct confidence limits for power, and is captured as a p-value function.  {Regardless of the test used to construct $H(\theta)$, Equation} (\ref{power cd}) {can be seen as a $g\{\theta\}=\beta^{-1}\beta\{\theta\}$ or {$g\{\beta\}=\beta^{-1}\{\beta\}$} link function to produce inference on power.  In terms of a Wald test for $\theta$ using an identity link, a Wald test for power using Equation} (\ref{power cd}) {would be $H(\beta)=1-\Phi([\hat{\theta}-\beta^{-1}\{\beta\}]/\hat{\text{se}})$.}  
Using the invariance property, $\hat{\beta}_{mle}=\beta(\hat{\theta}_{mle})$ is the maximum likelihood estimate for power, and in general $\hat{\beta}=\beta(\hat{\theta})$ is a plug-in estimate of power.  Such estimation is purely a function of the data and not of any population parameters, and so in this sense would be considered unconditional.  To better account for having estimated unknown nuisance parameters from external studies to estimate power, one could utilize a transformation of the power point estimate along with the delta method and conduct a t- or Wald test to ultimately construct $H(\beta)$, {under mild regularity conditions and standard asymptotics} (see Appendix \ref{delta method} for further mathematical considerations).  {Alternatively, without necessarily appealing to regularity conditions and standard asymptotics, one can derive or approximate the sampling distribution of the estimator for power and utilize its cumulative distribution function with restricted estimates of the nuisance parameters to construct $H(\beta)$.}  Of course, to make use of all available information during inference, one could construct the profile likelihood for power and conduct the likelihood ratio test.  
{When $H(\theta)$ has the appearance of a distribution function on the parameter space}, one can calculate what is otherwise considered a Bayesian quantity known as \textit{probability of success}, or \textit{assurance}, 
\begin{eqnarray}\label{pos}
\hat{\beta}_{pos}&=&\int \beta(\theta) \cdot dH(\theta)\\
&=&\int \beta \cdot dH(\beta).\label{pos2}
\end{eqnarray}
\\

{To the Bayesian, $H(\theta)$ is constructed using Bayes' theorem and is said to measure the experimenter's freely assigned belief about $\theta$ for the treatment under investigation, so that probability of success is not an estimate of the long-run probability of achieving end-of-study success, it is the belief about achieving end-of-study success.  A value of 0.5 represents complete uncertainty in belief or a lack of knowledge.  Probability of success is \textit{un}-conditional on $\theta$, but it \textit{does} depend on the freely assigned belief about $\theta$.  To the frequentist, there is a single true $\theta$ for the treatment  under investigation and} (\ref{pos}) {is the average of all possible hypotheses for power over the ex-post sampling probability in $H(\theta)$ (a p-value function).  
Equations  (\ref{pos}) and (\ref{pos2}) are unconditional in that they are not functions of $\theta$, and represent a point estimate of power.  Although consistent as an estimator, it is biased towards 0.5 since $\theta$ is a fixed quantity.  The uncertainty around having estimated power using $\hat{\beta}_{mle}$ and $\hat{\beta}_{pos}$ is not ignored, it is displayed in the p-value function for power.} 
\\

Probability of success is typically approximated through numerical integration by sampling from $H(
\theta)$.  However, once $H(\beta)$ is constructed as outlined above, probability of success can be easily approximated using a Riemann sum
\begin{eqnarray}
\hat{\beta}_{pos}&\approx&\frac{\sum \beta\cdot \Delta H(\beta)}{\sum \Delta H(\beta)}\nonumber\\
&=&\frac{\sum \beta(\theta)\cdot \Delta H(\theta)}{\sum \Delta H(\theta)}.\label{pos approximation}
\end{eqnarray}
This can be accomplished in a single data step and a call to Proc Means with a weight statement, and computes in a fraction of a second.  When considering two separate studies, e.g. phase 2 and phase 3 of a clinical development plan, probability of success can be defined as

\begin{eqnarray}
\hat{\beta}^{pos}_{2,3}=\int\beta_2(\theta)\beta_3(\theta)   \cdot    dH(\theta),\label{pos simplified theta}
\end{eqnarray}
where $\beta_2$ and $\beta_3$ are phase 2 and phase 3 power respectively.  This is easily approximated as in Equation (\ref{pos approximation}).  Reading Equation (\ref{pos simplified theta}) from left to right, for a given $\theta$, $\beta_2(\theta)\beta_3(\theta)$ is the power of succeeding in both phase 2 and phase 3, averaged over what is currently inferred about $\theta$.  In this quantity the truth does not change from phase 2 to phase 3, and probability of success is based solely on what is inferred now about $\theta$.  In a Bayesian framework, any unknown nuisance parameters would be treated as random variables.  This requires an additional layer of averaging when calculating probability of success, but typically has little impact on the result.  This integration of nuisance parameters would be analogous to the restricted estimation of nuisance parameters in the frequentist framework as is performed in the likelihood ratio and the score tests.

\section{Decision Making Across Pharmaceutical Development}\label{QDM}

\subsection{Decision Rules for End-of-Study Success}\label{success criteria}
Regardless of which paradigm one operates under, hypothesis testing is the very heart of quantitative decision making in pharmaceutical development.  
The null value to be tested in each phase depends not only on regulatory requirements, but also on what is clinically meaningful and commercially viable.  When showing a treatment effect over placebo or an active comparator, the null value need not be zero and the significance level need not be 0.05.  The example below uses confidence curves to visualize the success criteria in a phase 2 and 3 clinical development plan.     
\\

\textit{Example}:  A phase 2 and 3 development plan is being created for an asset to treat an immuno-inflammation disorder.  Phase 3 is planned as a non-inferiority study using a difference in proportions on a binary responder index.  The non-inferiority margin is set by the regulatory agency at $-0.12$, as is the one-sided significance level of 0.025.  Phase 2 is a dose finding study on a continuous endpoint.  This study also collects data on the responder index and includes a control arm to estimate the difference in proportions planned for phase 3.  A stricter non-inferiority margin of $-0.05$ is considered in phase 2, but since the sample size in phase 2 is typically smaller than in phase 3, a larger one-sided significance level of 0.20 is tolerated.  Based on a literature review the estimated response proportion for the comparator is 0.43 with N=1200. 
\\

{For each study let $X_{ctrl}\sim \text{Bin}(n_{ctrl},p_{ctrl})$ be the number of responders out of $n_{ctrl}$ subjects in the control group and $X_{active}\sim \text{Bin}(n_{active},p_{active})$ be the number of responders out of $n_{active}$ subjects in the active group, with $\theta=p_{active}-p_{ctrl}$, and $p_{ctrl},\theta\in\Theta$.}  {Then the corresponding likelihood function for each study is} $L(\theta,p_{ctrl}) \propto (p_{ctrl})^{x_{ctrl}}(1-p_{ctrl})^{n_{ctrl}-x_{ctrl}}(p_{ctrl}+\theta)^{x_{active}}(1-p_{ctrl}-\theta)^{n_{active}-x_{active}}$.
Figure \ref{binary success} uses confidence curves resulting from likelihood ratio tests on the population-level difference in proportions $\theta$ to demonstrate what the minimum phase 2 and phase 3 success criteria defined above look like in terms of a particular experimental result.  Nearly identical confidence curves can be produced by conducting Wald tests using identity links.  
The left panel is based on N=90 subjects per arm with an estimated response rate of 0.43 on the control arm, and an estimated difference in proportions of 0.01 (critical effect size).  This particular experimental result produces a p-value just under 0.20 when testing against the $-0.05$ non-inferiority margin, $H_0$: $\theta\le -0.05$.  This ex-post sampling probability forms the level of confidence that $\theta$ is less than or equal to $-0.05$.  That is, one must be at least 80\% confident that the true difference in proportions is greater than $-0.05$ in order to succeed in phase 2.  
As evidenced by the left panel in the figure below, declaring success for this experimental result is nearly equivalent to a test about the $-0.12$ non-inferiority margin at the 0.025 significance level.  The right panel is based on N=365 subjects per arm and an estimated difference in proportions of $-0.05$. This results in a p-value just under 0.025 when testing $H_0$: $\theta\le -0.12$, or equivalently, one must be at least 97.5\% confident that the true difference in proportions is greater than $-0.12$.  The phase 2 null hypothesis was chosen as the value at which phase 3 power is 50\%.  This will be seen more clearly in Section \ref{conditioning on phase 2}.  See Appendix \ref{appendix LR test} for the mathematical considerations regarding these decision rules and Appendix \ref{sas code} for the corresponding SAS code.  Such notation is suppressed here for ease of reading.


\begin{figure}[H]
	\centering
		{\includegraphics[trim={1cm 18.75cm 0 0},clip, height =2.2in]{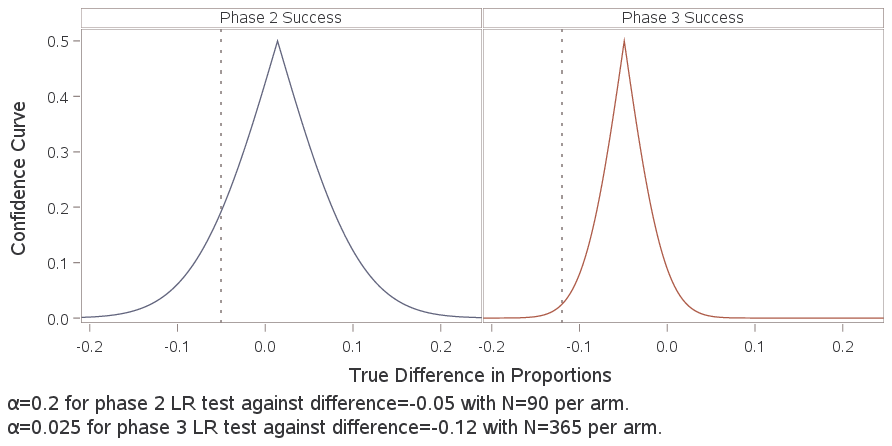}}
	\caption{\small{Phase 2 likelihood ratio test of $H_0$: $\theta\le-0.05$ with N=90 per arm at $\alpha$=0.2.  Phase 3 likelihood ratio test of $H_0$: $\theta\le-0.12$ with N=365 per arm at $\alpha$=0.025.} }
	\label{binary success}
\end{figure}

While it is important to have a clear definition of technical success before conducting a trial, Figure \ref{binary success} makes it clear there is nothing materially different between a p-value of 0.024 and 0.026, or 0.19 and 0.21 and so on.  This allows for flexibility in decision making and reminds us that no hypothesis is proven false with a single small p-value, nor is it proven true with a large one.  All we can do is provide the weight of the evidence.  This resonates with the American Statistical Association (ASA) statement on statistical significance and the p-value (\citeauthor{wasserstein2016} 2016).  It also reflects the original intentions of Fisher's statistical significance and inductive reasoning using a frequentist interpretation of probability (\citeauthor{lehmann1993} 1993).  
Equally important as the end-of-study success rule is the power of achieving it.  Both of these factor into the Go/No-Go decision and it is not enough to provide a point estimate of power.  One must also perform inference on power.

\subsection{Priors, Power, and Probability of Success}\label{priors power and pos}
\subsubsection{Elicitation}\label{prior elicitation}
Expert opinion can be used to perform inference on the power of a future study when no historical data is available (\citeauthor{european2014} 2014).  Many times expert opinion is elicited through a ``chips-in-bins" activity to construct a distribution estimate of the true treatment effect (\citeauthor{oakley2010} 2010).  This of course is inadmissible as scientific evidence, but allows the Bayesian to explore belief probabilities and allows the frequentist to consider inference based on hypothetical experimental evidence.  The available knowledge and information can be seen as exchangeable virtual data, and each expert considers all possible point estimates that data like this could give rise to, essentially bootstrapping the sampling distribution of the estimator (\citeauthor{xie2013incorporating} 2013).  These bootstrapped sampling distributions are then averaged in some way to form a single distribution.  If the experts were all bootstrapping from the same information their distributions would be nearly indistinguishable, but this is rarely the case.  The heterogeneity between the experts' distributions suggests an extra layer of bootstrap sampling.  Each expert's perspective represents a bootstrapped sample of the available information, from which they bootstrap repeatedly to form their distribution.  This explains the heterogeneity, and in theory the heterogeneity should be ``averaged out" when these distributions are combined.  The combined sampling distribution itself may be considered an approximate p-value function, but can also be used to conduct a hypothesis test.    See Appendix \ref{construction of densities} for the connection between an estimated sampling distribution and a p-value function.  
\\

\textit{Example continued}:  Six experts were assembled to elicit a distribution estimate for the difference in proportions of the responder index in the target patient population.  After a briefing on the literature to date all six experts' distributions were averaged to form a single estimated sampling distribution with a mean of $-0.02$.  This mean was used as the maximum likelihood point estimate for a likelihood ratio test of the difference in proportions based on N=350 on the investigational product, a 0.43 response rate in the control arm with N=1200, and used to form a confidence curve.  The virtual or effective sample size was determined by the variance of the combined sampling distribution 
and the literature review (see Appendix \ref{determining effective sample size}).  

\begin{figure}[H]
	\centering
	\includegraphics[trim={0cm 18.75cm 0 0}, clip, height = 2.25in]{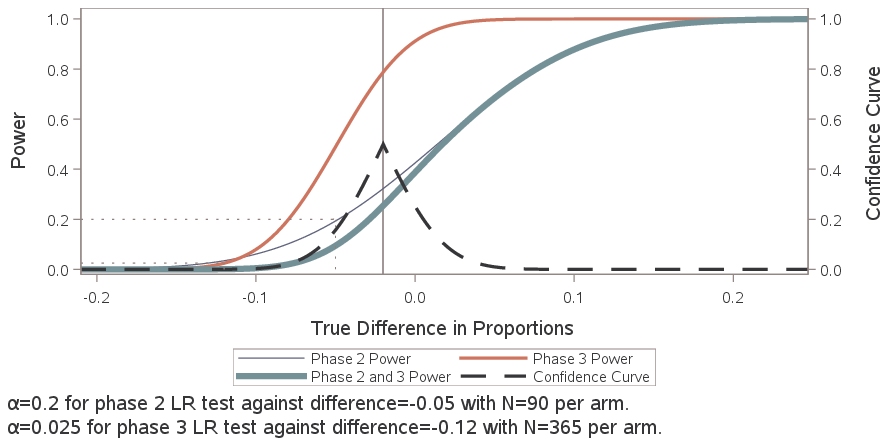}
\caption{\small{Phase 2 power curve testing $H_0$: $\theta\le-0.05$ with N=90 per arm at $\alpha$=0.2.  Phase 3 power curve testing $H_0$: $\theta\le-0.12$ with N=365 per arm at  $\alpha$=0.025.  Confidence curve for $\theta$ based on historical data and expert opinion. }} 
	\label{power figure}
\end{figure}
\vspace{2mm}

Figure \ref{power figure} shows the power curves for the success criteria outlined in Section \ref{success criteria}, 
the combined power curve (product) for success in both phase 2 and phase 3, and the elicited confidence curve for the difference in proportions described above.  The power curves in Figure \ref{power figure} are constructed while estimating the unknown true population-level response rate on the control therapy as 0.43 based on the literature review, approximated using the upper p-value function from a likelihood ratio test.  This approximation is nearly equivalent to using an upper p-value function from a Wald test.  (See Appendix \ref{extrapolation} for how to extrapolate the estimated power curve between endpoints or control groups across phases of development.) 
\\

Figure \ref{power figure cd} shows the resulting confidence curves for power using Equation (\ref{power cd}) and probability of success calculations using (\ref{pos approximation}) and (\ref{pos simplified theta}) based on the elicitation and literature review shown in Figure \ref{power figure}.  Figures \ref{power figure} and \ref{power figure cd} suggest a larger sample size in phase 2 would be warranted to increase the maximum likelihood and probability of success estimates for power in phase 2 and overall.  If 80\% or 90\% power is desired in the phase 3 study its sample size would need to be increased as well.  However, these statements ignore the inference in the confidence curves (see Figure \ref{power_plot}).  The bias of $\hat{\beta}_{PoS}$ makes it a useful summary measure since a relatively high or low value indicates the inference is centered near high or low values of power respectively, but this still does not provide a complete picture.  For instance, had the elicited confidence distribution been wider and shifted to the right probability of success would increase at most sample sizes, but this produces a U-shaped confidence density around power (\citeauthor{rufibach2016bayesian} 2016) (see Appendix \ref{additional figures}).  Since the confidence curve displays the same inference and is always concave it may be a better choice than the confidence density as in Figure \ref{power figure cd} for displaying inference on power.  
Of course the elicitation is merely hypothetical evidence.  What matters more is inference based on real data.  For this, one will need to conduct the phase 2 study.




\begin{figure}[H]
	\centering
	\includegraphics[trim={1cm 16.75cm 0 0}, height = 3in]{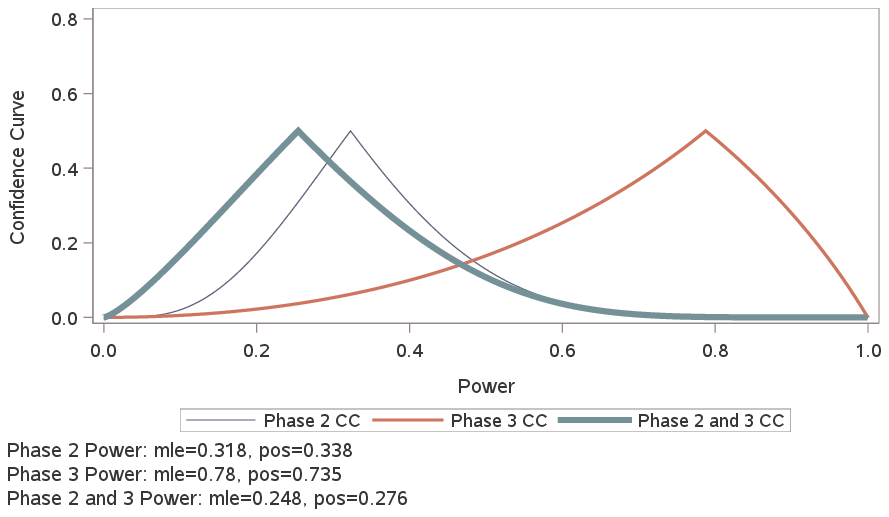} 
	\caption{\small{Solid lines depict resulting confidence curves for power in phase 2, phase 3, and overall based on the elicitation.  Peaks correspond to maximum likelihood estimates of power.} }
	\label{power figure cd}
\end{figure}



\begin{figure}[H]
	\centering
	\includegraphics[trim={1cm 18.7205cm 0 0}, clip, height = 2.25in]{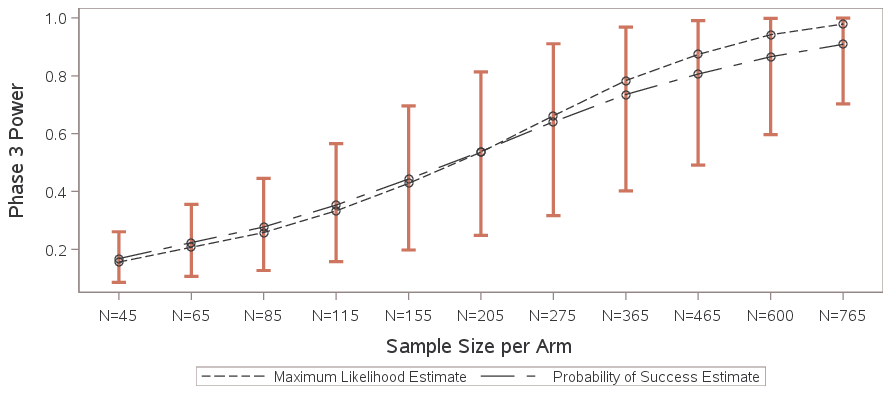}
	\caption{\small{Estimated phase 3 power when testing $H_0$: $\theta\le-0.12$ at $\alpha$=0.025 at various sample sizes with 80\% two-sided confidence limits based on the elicitation. }} 
\label{power_plot}
\end{figure}

\subsubsection{Conditioning on Phase 2 Success}\label{conditioning on phase 2}
If one is satisfied with the inference on phase 3 power given minimal success in phase 2, one would be satisfied for any other successful phase 2 result.  Recall the estimated phase 2 power curve was approximated using a p-value function.  The confidence curve depicting minimum success in phase 2 is simply a re-expression of this p-value function.  This is depicted in Figure \ref{power figure cond} and shows that the phase 2 decision rule from Figure \ref{binary success} produces inference around high values of phase 3 power, but still assumes some risk.  While the maximum likelihood and probability of success point estimates for phase 3 power are 95.9\% and 78.1\% respectively, one can claim with only 80\% confidence that the power of the phase 3 study is no less than 50\% given minimal success in phase 2 (p-value = 0.2 testing $H_0$: $\beta_3(\theta) \le 0.5$).  In our view, ensuring phase 3 power is no worse than a coin toss conditional on passing phase 2 is a good rule of thumb.  If stronger inference on phase 3 power is desired given minimal success in phase 2, one could simply increase the phase 3 sample size.  Alternatively, one could adjust the phase 2 significance level and null hypothesis, and select the phase 2 sample size based either on an acceptable phase 2 minimum clinically important difference, or on an acceptable observable critical effect size.  
 Once the phase 2 study results are available, 
two-sided confidence limits for phase 3 power can be provided alongside the maximum likelihood 
point estimate.  Conversely, the p-value testing $H_0$: $\beta_3(\theta) \le 0.5$ or the level of confidence for which phase 3 power is greater than 50\% can be provided alongside the point estimate.  Figures \ref{power figure cd} and \ref{power_plot} could be reproduced using phase 2 inference instead of the elicitation.  As mentioned in the introduction, if inference on power is ignored the decision maker may otherwise be indifferent and unwittingly exposed to risk when choosing programs to progress to phase 3 based on point estimates of power.  If \textit{probability of success} or \textit{assurance} is utilized as an estimate of the probability of achieving end-of-study success, we recommend not presenting it as an unconditional quantity that transcends power and does not require inference.  If \textit{probability of success} or \textit{assurance} is utilized as the unconditional confidence level of a prediction interval, 
we recommend not presenting it as the probability of achieving end-of-study success despite its namesake.  See Section \ref{sim section} for further discussion on interpreting prediction intervals.  



\begin{figure}[H]
	\centering
	\includegraphics[trim={0cm 18.74cm 0 0}, clip, height = 2.25in]{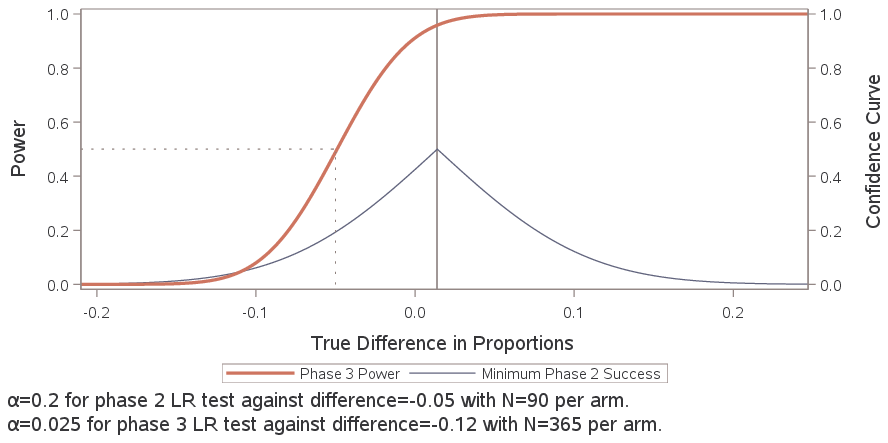}
	\caption{\small{ Phase 3 power curve testing $H_0$: $\theta\le-0.12$ with N=365 per arm at  $\alpha$=0.025.  Confidence curve for $\theta$ from the approximate phase 2 power curve testing $H_0$: $\theta\le-0.05$ with N=90 per arm at $\alpha$=0.2.} }
	\label{power figure cond}
\end{figure}

The inference above is conditional on minimal success in phase 2 alone.  One might also be interested in performing inference on phase 3 power that incorporates the elicited distribution estimate, though this should not weigh too heavily on decision making.  Often the phase 3 probability of success calculation is estimated through simulation while treating the elicited $h(\theta)$ as a probability distribution for $\theta$, and is conditioned on those Monte Carlo runs where the phase 2 success criteria is met.  This subsetting amounts to multiplying the phase 2 power curve by the elicited $h(\theta)$ and normalizing, $\beta_2(\theta)\cdot h(\theta)/\int\beta_2(\theta)\cdot h(\theta)d\theta$.  
In a Bayesian framework $\theta$ is considered ``random" and this density is conditional on the elicited $h(\theta)$ and on passing phase 2, but without conditioning on a particular value of $\theta$ nor a particular phase 2 result.  This density, sometimes referred to as a pre-posterior, and the phase 3 power curve produce the \textit{conditional probability of success}, or \textit{conditional assurance} (\citeauthor{temple2021} 2021), estimate of power.  This is similar to though not exactly the same as multiplying the elicited $H(\theta)$ by the approximate estimated phase 2 power curve (minimum end-of-study success upper p-value function) and differentiating, $d\big( H(\theta)\cdot\beta_2(\theta)\big)/d\theta$.  This same inference can be displayed as a confidence curve.  See curve (iii) in Figure \ref{mult vs conv} below.   The fixed $\theta$ interpretation of this curve is the upper-tailed probability of observing a result as or more extreme than the elicited test statistic \textit{and} a result as or more extreme than the critical effect size in phase 2, given hypotheses of the form $H_0$: $\theta\le\theta_0$.  This same curve depicts lower-tailed \textit{or} probability statements testing hypotheses of the form $H_0$: $\theta\ge\theta_0$.  In this inference the elicited point estimate and the phase 2 point estimate are treated as separate observations.  The median of this p-value function (two-sided p-value = 1) can be used as a point estimate for $\theta$ and to form a point estimate for phase 3 power.  Alternatively, one could convolve the approximate estimated phase 2 power curve (minimum end-of-study success upper p-value function) with the elicited $H(\theta)$ using Equation (\ref{conf_dist_2}) to form the updated p-value function for the treatment effect.  See curve (iv) in Figure \ref{mult vs conv} below.  This convolution treats the elicitation and the phase 2 study as a single larger study.  
See Appendix \ref{additional figures3} for additional figures.  
\\

This process of performing inference on power can be extended to include multiple phase 2 power curves, with or without the elicited $H(\theta)$, and sequentially updating the p-value function for the treatment effect by multiplying or convolving the p-value functions as described above.  For example, inference on phase 2a, phase 2b, phase 3, and overall power conditional on passing a pilot study; inference on phase 2b, phase 3, and overall power conditional on passing the pilot and phase 2a studies; inference on phase 3 power conditional on passing the pilot and phase 2a and 2b studies.  If one is dissatisfied with the inference on phase 3 power after the phase 2 study results are observed, one could consider increasing the phase 3 sample size.  This will steepen the phase 3 power curve relative to the phase 3 null hypothesis by lowering the critical effect size, and improve the inference on phase 3 success.  Figures \ref{power figure cd} and \ref{power_plot} could be reproduced using phase 2 inference instead of the elicitation.  Of course one could also consider conducting an additional phase 2 study and multiply or convolve the results with the other observed phase 2 p-value functions.  The observed phase 2 results could be used to update the estimated phase 3 power curve by combining estimates of population-level nuisance parameters, and inference on phase 3 power could be constructed using the delta method.


\begin{figure}[H]
	\centering
\includegraphics[trim={0cm 17.6cm 0 0}, clip,height = 2.7in]{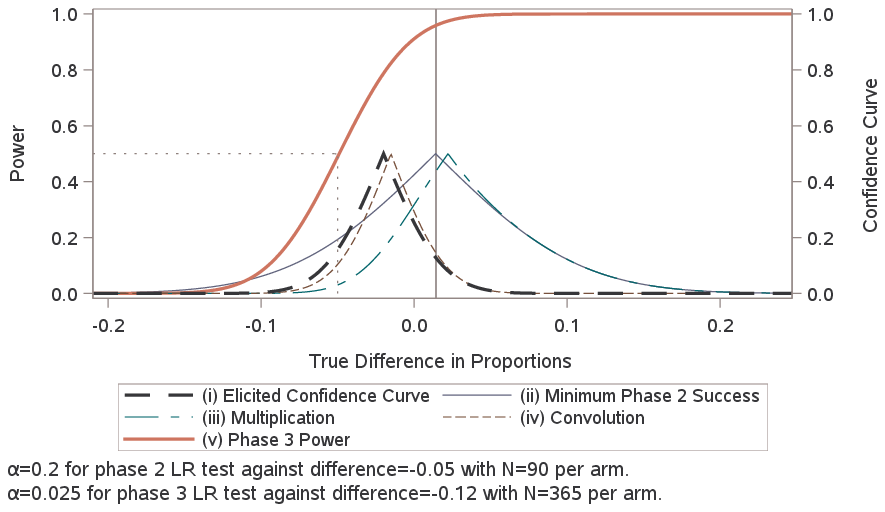}
	\caption{\small{(i) Elicited confidence curve.  (ii) Confidence curve for $\theta$ from the approximate phase 2 power curve  testing $H_0$: $\theta\le-0.05$ with N=90 per arm at $\alpha$=0.2.  (iii) Multiplication of elicited $H(\theta)$ and phase 2 power curve, displayed as a confidence curve.   (iv) Convolution of elicited $H(\theta)$ and approximate phase 2 power curve, displayed as a confidence curve.  (v) Phase 3 power curve testing $H_0$: $\theta\le-0.12$ with N=365 per arm at  $\alpha$=0.025. } }
\label{mult vs conv}
\end{figure}

\section{Simulation Study}\label{sim section}
Here we consider a simulation scenario that closely resembles Figure  \ref{power figure cond} to investigate the performance of decision rules based on point estimates and confidence intervals for power.  Without including any external or elicited data a phase 2 sample of size $N=90$ per arm is simulated and used to estimate the phase 3 power curve with N=365 per arm investigating a difference in proportions $\theta$ by testing $H_0$: $\theta \le -0.12$ at the 0.025 significance level using a likelihood ratio test.  Operating characteristics of decision rules for progression into phase 3 based on the maximum likelihood and probability of success estimates of power are presented in Table \ref{sim table}, as well as a decision rule based on a one-sided 80\% confidence interval for power using the approach corresponding to Equation (\ref{power cd}).  Three treatment effect scenarios are investigated: $\theta=-0.12$, $\theta=-0.05$, and $\theta=0$.  In each scenario the unknown true population-level control therapy response rate is 0.43.  The decision rule labeled `PoS$\ge 0.60$' represents a Go decision into phase 3 if the probability of success estimate of power is greater than or equal to 0.60.  Likewise for `PoS$\ge 0.75$' and `PoS$\ge 0.80$'.  The rule labeled `MLE$\ge 0.80$' represents a Go decision into phase 3 if the maximum likelihood estimate of power is greater than or equal to 0.80, and the rule labeled `80\% Conf. $\beta_3 > 0.50$' represents a Go decision into phase 3 if the test $H_0$: $\beta_3 \le 0.50$ is significant at the 0.20 level.  
\\

The two-sided 60\% confidence interval for phase 3 power based on phase 2 results using the approach corresponding to Equation (\ref{power cd}) covered 60.4\%, 59.2\%, and 59.6\% of the time when the true power was 0.025, 0.50, and 0.91, respectively.  Comparatively, the two-sided 60\% confidence interval based on a Wald test using the delta method with a $g\{\cdot\}=\Phi^{-1}\{\cdot\}$ transformation of the maximum likelihood estimate of power covered 60.5\%, 59.2\%, and 59.6\% of the time.  Table \ref{sim table} shows that over 10,000 simulations the decision rule based on the one-sided 80\% confidence interval made a Go decision into phase three 19.3\% of the time if $\theta=-0.05$ and $\beta_3=0.50$.  This corresponds with the definition of the Go rule.  For the same simulation scenario the `PoS$\ge 0.60$', `PoS$\ge 0.75$', and `PoS$\ge 0.80$' decision rules made a Go decision into phase three 34.0\%, 15.2\%, and 10.4\% of the time respectively.  These results demonstrate that it is not immediately obvious how the probability of success estimate corresponds to the operating characteristics of a decision rule in relation to the true value of power.  
Compared to the decision rule based on the maximum likelihood estimate, the confidence interval rule works to guard against making a Go decision if the true power is low.  This of course is the intention behind the rules using the probability of success estimate, but the confidence interval rule does so with easily understood and controllable operating characteristics that define the rule itself.  Investigating the operating characteristics of several probability of success decision rules via simulation and selecting the rule with desirable characteristics is no different in principle from forming a confidence interval rule.  One could view a probability of success decision rule as the confidence level of a prediction interval for the phase 3 test statistic, which does have easily understood operating characteristics, e.g. a one-sided 75\% prediction interval will correctly predict the phase 3 result 75\% of the time regardless of the unknown fixed phase 3 power.  This would correspond to a `PoS$\ge0.75$' decision rule, but this confidence level is a statement about both the phase 2 and phase 3 sampling variability and it is impossible to tease this apart.  
In contrast, for inference on phase 3 power the confidence level relates only to phase 2 sampling variability, and hypotheses for phase 3 power pertain only to phase 3 uncertainty.  This makes inference on power much more meaningful and easier to interpret, which should lead to better decision making compared to predictive inference on success.  

\begin{table}[H]
	\caption{Simulation Results}
	
	\begin{center} 
		
		\begin{tabular}{  l r r r r r  }
	 & & & & & \\
	Unknown True  & & & & & $80\%$ Conf. \\
	 Phase 3 power & PoS $\ge 0.60$  &PoS $\ge 0.75$ & PoS $\ge 0.80$  & MLE $\ge 0.80$ & $\beta_3>0.50$\\
			 	 &          &      &       &  & \\
			\cline{1-1} \cline{2-2} \cline{3-3} \cline{4-4}  \cline{5-5}  \cline{6-6}    
			     ~ &     ~ & ~ &~&~&~\\
			  $\beta_3(\theta=-0.12)=0.025$    &0.091        & 0.023	& 0.015&0.079 & 0.034\\
			       $\beta_3(\theta=-0.05)=0.50$    & 0.340    &0.152	  &0.104  &0.329	&0.193\\
			 $\beta_3(\theta=0)=0.91$  &0.599    & 0.366 &0.263	 &0.606&0.428\\
			      &  &	&  & &\\
	\cline{1-1} \cline{2-2} \cline{3-3} \cline{4-4}  \cline{5-5}  \cline{6-6}  
		\end{tabular}
	\end{center}
	\small{\space\space\space \hspace{8mm} Operating characteristics of decision rules over 10,000 simulations.}
\label{sim table}
\end{table}

The results of the decision rule based on the confidence interval for power in Table \ref{sim table} should be clear from inspecting Figure \ref{power figure} since the estimated power curves in this figure match the unknown true power curves in the simulation study.  Considering the results from Table \ref{sim table}, the confidence interval rule produces a significant result 3.4\% of the time when testing $H_0$: $\beta_3 \le 0.50$ if $\theta=-0.12$ and $\beta_3=0.025$.  The phase 3 power curve in Figure \ref{power figure} evaluated at $\theta=-0.12$ is 0.025 and the phase 2 power curve is approximately equal to 0.034.  Similarly, considering again the results from Table \ref{sim table} the confidence interval rule produces a significant result 42.8\% of the time when testing $H_0$: $\beta_3 \le 0.50$ if $\theta=0$ and $\beta_3=0.91$.  The phase 3 power curve in Figure \ref{power figure} evaluated at $\theta=0$ is 0.91 and the phase 2 power curve is approximately equal to 0.428.  Increasing the phase 2 sample size will improve upon this 0.428 probability of making a Go decision if the true phase 3 power is 0.91, without altering the performance of the rule if the true phase 3 power is 0.50.  If different operating characteristics under $H_0$: $\beta_3 \le 0.50$ are desired, or if a different hypothesis is of interest, one can construct a different rule.    
\\

In practice we will not know which point on our estimated power curve corresponds most closely with the true value of power, and we will not actually repeat each experiment 10,000 times; however, the frequency probabilities concerning the experiment contained in the p-value function for power as a function of the hypothesis and the observed data provide the experimenter confidence when performing inference and making a decision.  As alluded to in Section \ref{success criteria}, decision making should be flexible.  There may be an experimental result with a small p-value for which it should be decided not to progress into phase 3 based on, say, market data, safety data, etc., and vice versa.  Ultimately it is up to the experimenter to make an informed decision, and the confidence provided by the p-value is part of that decision.  

\begin{figure}[H]
	\centering
	\includegraphics[trim={0cm 16cm 0 0}, clip, height = 2.75in]{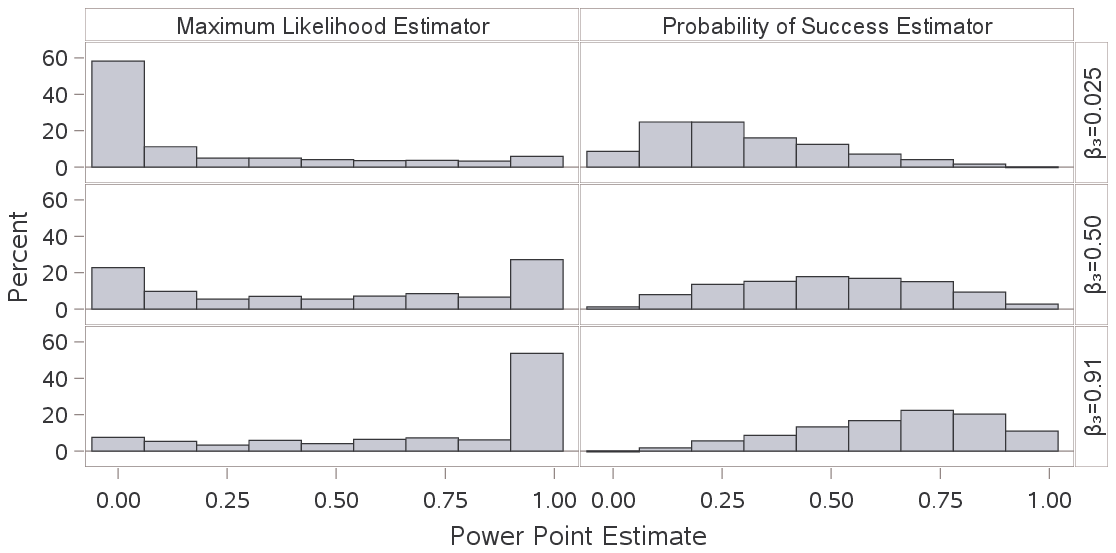}
	\caption{\small{Sampling distributions of the maximum likelihood and probability of success estimators of power over 10,000 simulations.} }
	\label{sim plot}
\end{figure}

Figure \ref{sim plot} shows the sampling distributions of the maximum likelihood and probability of success estimators of power over the 10,000 simulations.  In repeated sampling the probability of success estimator tends to produce a value not far from 0.50, whether the true power of the phase 3 study is 0.91, 0.50, or 0.025.  In this setting the maximum likelihood estimator of power is median-unbiased, producing estimates centered around the true value of power.  The sampling distribution of the $\Phi^{-1}$ transformed maximum likelihood estimator of power is shown in Appendix \ref{sim figure}.  The inverse cumulative distribution function of the standard normal distribution works incredibly well at stabilizing the variance and producing an approximately normal sampling distribution.  This allows for constructing a p-value function for power using a Wald test with the delta method instead of Equation (\ref{power cd}).

\section{Closing Remarks}
The {p-value function} is a remarkable visual tool for displaying quantitative decision rules and study results, and can even be used to display inference on power.  The Bayesian quantity \textit{probability of success} or \textit{assurance}, whether viewed as the confidence level of a prediction interval, the result of a biased estimator of power, or a philosophical value, may not be the primary quantity of interest for decision making in drug development.  {Although our demonstrations focused on an exponential family model with routine asymptotic tests, the construction of a p-value function for power is not limited to this setting.  A natural extension of our work would be to perform inference on power by jointly modeling correlated endpoints, and perhaps even constructing a confidence region for power.}  
While not demonstrated herein, confidence densities and confidence curves can also be used for conducting interim analyses.  Stopping rules for early efficacy based on p-values would be displayed similarly to Figure \ref{binary success} using the data at interim, while stopping rules for futility based on inference of end-of-study power given the data at interim would resemble Figure \ref{mult vs conv} with the p-value function for the treatment effect determined, at least in part, by the interim data.  

\pagebreak
\section*{Data Sharing}
Data sharing is not applicable to this article as no new data were created or analyzed in this study.

	\bibliographystyle{chicago}
	\bibliography{bib}

	\pagebreak

	\appendix

\section{Definitions}\label{definitions}
\subsection{Definition of a Confidence Interval}\label{CI definition}
From \cite{casella2002}, the inference in a set estimation problem is the statement that `$\theta \in C$,' where $C \subset \Theta$ and $C=C(\boldsymbol{x})$ is a set determined by the value of the data $\boldsymbol{X}=\boldsymbol{x}$ observed.  $C \subset \Theta$ is usually taken to be an interval, and $C(\boldsymbol{X})$ is its estimator, a random variable.  The coverage probability, $P_{\theta}(\theta \in C(\boldsymbol{X}))$, is a probability statement referring to the random set $C(\boldsymbol{X})$ since $\theta$ is an unknown fixed quantity.  
	
\subsection{Definition of a Confidence Distribution}\label{definition}
From \cite{xie2013incorporating}, a function $H_n(\cdot)$ on $\mathcal{X} \times \Theta \rightarrow [0,1]$ is called a confidence distribution function for a parameter $\theta$ if, R1) For each given $\boldsymbol{x} \in \mathcal{X}$, $H_n(\cdot)$ is a cumulative distribution function on $\Theta$; R2) At the true parameter value $\theta=\theta_0$, $H_n(\theta_0)\equiv H_n(\boldsymbol{x},\theta_0)$, as a function of the sample $\boldsymbol{x}$, follows the uniform distribution $U[0,1]$.  $H_n(\cdot)$ is an asymptotic confidence distribution if the $U[0,1]$ requirement is true only asymptotically, and the continuity requirement on $H_n(\cdot)$ is dropped.

\section{Mathematical Considerations}

\subsection{{Link Functions}}\label{link functions}
The Wald test is incredibly versatile, especially when incorporating a link function.  A link function can also be helpful with a score or likelihood ratio test when the referenced sampling distribution is approximate.  This is often used in the analysis of generalized linear models where $g\{\cdot\}$ is a log or logit transformation.  Careful selection of the link function can vastly improve the inference on a parameter.  For example, consider the setting where $X_1,...,X_n\sim N(\theta,1)$ and interest surrounds $\beta=-1/\theta$.  Using $\hat{\beta}=-1/\bar{x}$ and an identity link,  $H(\beta)=1-\Phi\big(\bar{x}\sqrt{n}[-1/\bar{x}-\beta]\big)$ is a reasonable approximate solution since $\bar{X}\sqrt{n}(-1/\bar{X}-\beta)\overset{asymp}{\sim}N(0,1)$, so long as $\theta \ne 0$.  However, a $g\{\beta\}=1/\beta$ link function leads to  $\sqrt{n}(\bar{X}-\frac{-1}{\beta})\sim N(0,1)$, producing exact inference using $H(\beta)=1-\Phi\big(\sqrt{n}[\bar{x}-\frac{-1}{\beta}]\big)$.  As another example, consider the setting where we have two sets of normal samples from $N(\theta_1,1)$ and $N(\theta_2,1)$ respectively and interest surrounds $\beta=\theta_1/\theta_2$.  Using $\hat{\beta}=\bar{x}_1/\bar{x}_2$ and an identity link leads to approximate inference based on $(\bar{X}_1/\bar{X}_2-\beta)/\text{se} \overset{asymp}{\sim}N(0,1)$.  However, a $g\{\beta\}=\beta\cdot\bar{x}_2$ link function yields exact inference based on $(\bar{X}_1-\beta\cdot\bar{X}_2)/\text{se}\sim N(0,1)$.  Regardless of the test used to construct $H(\theta)$, Equation (\ref{power cd}) can be seen as a $g\{\theta\}=\beta^{-1}\beta\{\theta\}$ link function to produce inference on power.

\subsection{Determining Effective Sample Size}\label{determining effective sample size}
If a literature review and elicitation provides an estimated sampling distribution for the response proportion on control and the difference over control, the first two moments of these distributions can be used to determine the effective sample size for the active arm.
\begin{eqnarray}
\hat{\text{Var}}(\hat{p}_{active}-\hat{p}_{ctrl})&=&\frac{\hat{\sigma}^2_{active}}{n_{active}}+\frac{\hat{\sigma}^2_{ctrl}}{n_{ctrl}}\nonumber\\
\hat{\text{Var}}(\hat{p}_{active}-\hat{p}_{ctrl})&=&\frac{\hat{p}_{active}(1-\hat{p}_{active})}{n_{active}}+\frac{\hat{p}_{ctrl}(1-\hat{p}_{ctrl})}{n_{ctrl}}\nonumber\\
\hat{\text{Var}}(\hat{p}_{active}-\hat{p}_{ctrl})-\frac{\hat{p}_{ctrl}(1-\hat{p}_{ctrl})}{n_{ctrl}}&=&\frac{\hat{p}_{active}(1-\hat{p}_{active})}{n_{active}}\nonumber\\
n_{active}&=&\frac{\hat{p}_{active}(1-\hat{p}_{active})}{\hat{\text{Var}}(\hat{p}_{active}-\hat{p}_{ctrl})-\frac{\hat{p}_{ctrl}(1-\hat{p}_{ctrl})}{n_{ctrl}}}\nonumber
\end{eqnarray}

\subsection{Likelihood Ratio Test for Difference in Proportions}\label{appendix LR test}
This is a quick reference to performing the likelihood ratio test for a difference in proportions.  See Casella and Berger (2002) for complete instruction on the definition of symbols and how to construct a likelihood ratio hypothesis test.  
\\

Let $X_{ctrl}\sim \text{Bin}(n_{ctrl},p_{ctrl})$, $X_{active}\sim \text{Bin}(n_{active},p_{active})$, $\theta=p_{active}-p_{ctrl}$, and $p_{ctrl},\theta\in\Theta$.

\begin{eqnarray}
L(\theta,p_{ctrl}) &\propto& (p_{ctrl})^{x_{ctrl}}(1-p_{ctrl})^{n_{ctrl}-x_{ctrl}}(p_{ctrl}+\theta)^{x_{active}}(1-p_{ctrl}-\theta)^{n_{active}-x_{active}}\nonumber\\
\frac{\partial\ell(\theta,p_{ctrl})}{\partial p_{ctrl}}&=&\frac{x_{ctrl}}{p_{ctrl}}-\frac{n_{ctrl}-x_{ctrl}}{1-p_{ctrl}}+\frac{x_{active}}{p_{ctrl}+\theta}-\frac{n_{active}-x_{active}}{1-p_{ctrl}-\theta}\nonumber\\
\frac{\partial\ell(\theta,p_{ctrl})}{\partial \theta}&=&\frac{x_{active}}{p_{ctrl}+\theta}-\frac{n_{active}-x_{active}}{1-p_{ctrl}-\theta}\nonumber
\end{eqnarray}
$\underset{{p}_{ctrl},\theta\in\Theta}{sup}L(\theta,{p}_{ctrl})=L(\hat{\theta},\hat{p}_{ctrl})$ yields $\hat{p}_{ctrl}=x_{ctrl}/n_{ctrl}$ and $\hat{\theta}=x_{active}/n_{active}-x_{ctrl}/n_{ctrl}$. 
\\
\\
\\
 Under $H_0$: $\theta=\theta_0$, $\underset{{p}_{ctrl},\theta\in\Theta_0}{sup}L(\theta,{p}_{ctrl})=L(\theta_0,\hat{p}_{\theta_{0}}^{ctrl})$ where
\begin{eqnarray}
\frac{\partial\ell(\theta_0,p_{ctrl})}{\partial p_{ctrl}}&\overset{set}{=}&0\nonumber\\
\implies \hat{p}_{\theta_{0},1}^{ctrl}&=&\frac{x_{ctrl}+\frac{x_{active}\hat{p}_{ctrl}}{\hat{p}_{ctrl}+\theta_0}(1-\hat{p}_{ctrl})+\frac{x_{active}(1-\hat{p}_{ctrl})}{1-\hat{p}_{ctrl}-\theta_0}\hat{p}_{ctrl}}{n_{ctrl}+\frac{n_{active}(1-\hat{p}_{ctrl})}{1-\hat{p}_{ctrl}-\theta_0}}\nonumber\\
\hat{p}_{\theta_{0},k+1}^{ctrl}&=&\frac{x_{ctrl}+\frac{x_{active}\hat{p}_{\theta_{0},k}^{ctrl}}{\hat{p}_{\theta_{0},k}^{ctrl}+\theta_0}(1-\hat{p}_{\theta_{0},k}^{ctrl})+\frac{x_{active}(1-\hat{p}_{\theta_{0},k}^{ctrl})}{1-\hat{p}_{\theta_{0},k}^{ctrl}-\theta_0}\hat{p}_{\theta_{0},k}^{ctrl}}{n_{ctrl}+\frac{n_{active}(1-\hat{p}_{\theta_{0},k}^{ctrl})}{1-\hat{p}_{\theta_{0},k}^{ctrl}-\theta_0}}, k=1,2,...,K\nonumber
\end{eqnarray}
for $K$ sufficiently large to reach convergence. Estimating nuisance parameters under the restricted null space can also be accomplished in Proc Genmod by using the NOINT, OFFSET=, NOSCALE, and SCALE=  options in the MODEL statement.  In Proc Glimmix scale parameters are restricted using the HOLD= option in the PARMS statement.  Under mild regularity conditions the likelihood ratio test statistic,
\begin{eqnarray}
-2\text{log}LR(\boldsymbol{X},\theta_0)=-2\text{log}\Bigg(\frac{L(\theta_0,\hat{p}_{\theta_{0}}^{ctrl})}{L(\hat{\theta},\hat{p}_{ctrl})}\Bigg),\nonumber
\end{eqnarray}
follows an asymptotic chi-squared distribution with 1 degree of freedom, and significance at level $\alpha$ is achieved if $-2\text{log}LR(\boldsymbol{x},\theta_0)>\chi^2_{1,\alpha}$, the $1-\alpha$ percentile. The corresponding two-sided, equal-tailed $100(1-\alpha)\%$ confidence interval is given by $\{\theta:-2\text{log}LR(\boldsymbol{x},\theta)\le\chi^2_{1,\alpha}\}.$  The p-value function, confidence density, and confidence curve functionals for the test above are
\begin{eqnarray}
H(\theta_0,\boldsymbol{x})&=& \left\{ \begin{array}{cc}
\big[1-F_{\chi^2_1}\big(-2\text{log}LR(\boldsymbol{x},\theta_0)\big)\big]/2 & \text{if } \theta_0 \le \hat{\theta}(\boldsymbol{x}) \\
 &  \nonumber\\
 \big[1+F_{\chi^2_1}\big(-2\text{log}LR(\boldsymbol{x},\theta_0)\big)\big]/2  & \text{if } \theta_0 > \hat{\theta}(\boldsymbol{x}) \end{array}  \right.\nonumber\\
&&\nonumber\\
&&\nonumber\\
h(\theta_0,\boldsymbol{x})&=&\frac{dH(\theta_0,\boldsymbol{x})}{d\theta_0} \nonumber\\
&&\nonumber\\
&&\nonumber\\
C(\theta_0,\boldsymbol{x})&=& \left\{ \begin{array}{cc}
H(\theta_0,\boldsymbol{x}) & \text{if } \theta_0 \le \hat{\theta}(\boldsymbol{x}) \\
 &  \nonumber\\
 1-H(\theta_0,\boldsymbol{x})  & \text{if } \theta_0 \ge \hat{\theta}(\boldsymbol{x}). \end{array}  \right.\nonumber
\end{eqnarray} 
\\

The asymptotic result above in terms of the full likelihood is equivalently viewed as the profile likelihood ratio, 
\begin{eqnarray}
-2\text{log}\Bigg(\frac{L(\theta_0)}{L(\hat{\theta})}\Bigg)\overset{asymp}{\sim} \text{\raisebox{2pt}{$\mathlarger{\mathlarger{\chi}}$}}^2_1,\nonumber
\end{eqnarray}
where $L(\theta)=\underset{{p}_{ctrl}\in\Theta}{sup}L(\theta,{p}_{ctrl})=L(\theta,\hat{p}_{\theta}^{ctrl})$ as a function of $\theta$ and the observed data is the profile likelihood.  This replaces nuisance parameters with estimates calculated under the restricted parameter space, creating a one-dimensional likelihood.  $L(\hat{\theta})=L(\hat{\theta},\hat{p}_{ctrl})$ is the profile likelihood evaluated at $\hat{\theta}$ and $L(\theta_0)=L(\theta_0,\hat{p}_{\theta_{0}}^{ctrl})$ is the profile likelihood evaluated at $\theta_0$.  
\\

\subsection{Approximating Power using a P-value Function}\label{approximating power}
The proof that a p-value function can be used to approximate a power curve involves the continuous mapping theorem, convergence in probability, and convergence in distribution and is left to the reader as an exercise.  What follows is the intuition behind this approximation.  The upper p-value function has the appearance of a power curve for an upper-tailed test, and both depict sampling probability of the test statistic as a function of the unknown fixed true parameter value.  The p-value pertains to a specific experimental result and a single parameter unconditional on nuisance parameters, while power pertains to any statistically significant experimental result relative to a single research null hypothesis as a function of all unknown fixed parameters.  The p-value function is typically written as $H(\theta,\boldsymbol{x})$ to denote it as a function of both the parameter and the data.  This dependence on the data will enter through parameter estimates that are functions of the sufficient statistics, and so $H(\theta,\boldsymbol{x})$ can be expressed as $H(\theta,\hat{\theta},\hat{p}_{ctrl})$, where in our example $\hat{p}_{ctrl}$ is the point estimate for the population-level control therapy response rate $p_{ctrl}$, and $\hat{\theta}$ is the point estimate for the population-level difference in proportions $\theta$.  With a simple change of variables the p-values can be used to approximate power.  That is, if we consider an ex-ante experimental result where $\hat{p}_{ctrl}$ is exactly equal to $p_{ctrl}$, and $\hat{\theta}_{}$ equals the critical effect size $\hat{\theta}_{ces}$ for a research hypothesis of interest $\theta_0$, then $H(\theta,\hat{\theta}=\hat{\theta}_{ces},\hat{p}_{ctrl}=p_{ctrl})$ is a function of both $\theta$ and $p_{ctrl}$ and is approximately equal to the power of the test, $\beta(\theta,p_{ctrl})$.  When evaluated at $\theta=\theta_0$, $H(\theta=\theta_0,\hat{\theta}=\hat{\theta}_{ces},\hat{p}_{ctrl}=p_{ctrl})$ equals $\alpha$, the desired type I error rate of the test.  When evaluated at any other value of $\theta$, $H(\theta,\hat{\theta}=\hat{\theta}_{ces},\hat{p}_{ctrl}=p_{ctrl})\approx \beta(\theta,p_{ctrl})$.  
This same approach can be used to approximate the power of a lower-tailed test using a lower p-value function, denoted here as $H^{-}(\theta)$.  Since the approximate expression for power is a function of $\theta$ and $p_{ctrl}$, replacing $p_{ctrl}$ with a point estimate from an external study produces an estimated power curve as a function of $\theta$.  By replacing $\theta$ with a point estimate from an external study as well, the delta method can be employed to construct p-values and confidence intervals for hypotheses around power.  

\subsection{Delta Method for Inference on Power}\label{delta method}
Taylor Series
\begin{eqnarray}
g\big\{\beta\big(\hat{\theta}(\boldsymbol{X}),\hat{p}_{ctrl}(\boldsymbol{X})\big)\big\}&\approx& g\{\beta({\theta},{p}_{ctrl})\} + \frac{\partial g\{\beta({\theta},{p}_{ctrl})\}}{\partial \theta}\cdot\big(\hat{\theta}(\boldsymbol{X})-\theta\big) \nonumber\\
&&\hspace{21.5mm} + \frac{\partial g\{\beta({\theta},{p}_{ctrl})\}}{\partial p_{ctrl}}\cdot\big(\hat{p}_{ctrl}(\boldsymbol{X})-p_{ctrl}\big)\nonumber\\
&&\nonumber\\
&&\nonumber
\end{eqnarray}
Asymptotic Variance 
\begin{eqnarray}
{\text{Var}}\Big[g\big\{\beta\big(\hat{\theta}(\boldsymbol{X}),\hat{p}_{ctrl}(\boldsymbol{X})\big)\big\}\Big]&\approx&\hspace{4mm}\Bigg[\frac{\partial g\{\beta({\theta},{p}_{ctrl})\}}{\partial \theta}  \Bigg]^2 \cdot{\text{Var}}[\hat{\theta}(\boldsymbol{X})] \nonumber\\
&&\hspace{1.5mm}+ \Bigg[\frac{\partial g\{\beta({\theta},{p}_{ctrl})\}}{\partial p_{ctrl}}  \Bigg]^2\cdot {\text{Var}}[\hat{p}_{ctrl}(\boldsymbol{X})] \nonumber\\
&&+ 2\Bigg[\frac{\partial g\{\beta({\theta},{p}_{ctrl})\}}{\partial \theta}  \Bigg] \cdot \Bigg[\frac{\partial g\{\beta({\theta},{p}_{ctrl})\}}{\partial p_{ctrl}}  \Bigg] \cdot {\text{Cov}}[\hat{\theta}(\boldsymbol{X}),\hat{p}_{ctrl}(\boldsymbol{X})] \nonumber\\
&&\nonumber\\
&&\nonumber
\end{eqnarray}
Wald Confidence Interval for Power
\begin{eqnarray}
g^{-1}\Big[g\{\beta(\hat{\theta},\hat{p}_{ctrl})\}\pm z_{1-\alpha/2}\cdot\hat{\text{se}}\Big] \nonumber\\
&&\nonumber\\
&&\nonumber
\end{eqnarray}
Wald p-value testing $H_0$: $\beta \le \beta_0$
\begin{eqnarray}
H(\beta_0,\boldsymbol{x})=1-\Phi\Bigg(\frac{g\{\beta(\hat{\theta},\hat{p}_{ctrl})\}-g\{\beta_0\} }{\hat{\text{se}}} \Bigg)\nonumber\\
&&\nonumber\\
&&\nonumber
\end{eqnarray}
$\beta(\theta,p_{ctrl})$ is the unknown true power of a future study investigating a difference in proportions.  $\hat{p}_{ctrl}(\boldsymbol{X})$ is an estimator from an external study for the population-level response rate for the control therapy, $p_{ctrl}$.  $\hat{\theta}(\boldsymbol{X})$ is an estimator from an external study for the population-level difference in proportions between the experimental and control therapies, $\theta$.  $\beta\big(\hat{\theta}(\boldsymbol{X}),\hat{p}_{ctrl}(\boldsymbol{X})\big)$ is the corresponding estimator for power, and $g\{\cdot\}$ is a variance-stabilizing transformation that yields a normally distributed sampling distribution.  The Taylor series approximation is used to construct the asymptotic variance of the estimator for power.  Once the external data are observed the partial derivatives in the asymptotic variance can be solved numerically using parameter estimates, and ${\text{Var}}[\hat{\theta}(\boldsymbol{X})]$, ${\text{Var}}[\hat{p}_{ctrl}(\boldsymbol{X})]$, and ${\text{Cov}}[\hat{\theta}(\boldsymbol{X}),\hat{p}_{ctrl}(\boldsymbol{X})]$ can be replaced with model-based or sandwich estimates.  This produces an asymptotic variance estimate.  The estimated standard error $\hat{\text{se}}$ is the square root of the asymptotic variance estimate.
\\

\subsection{Transformed Power Estimator}\label{sim figure}

\begin{figure}[H]
	\centering
	\includegraphics[trim={0cm 15cm 0 0}, clip, height = 3.05in]{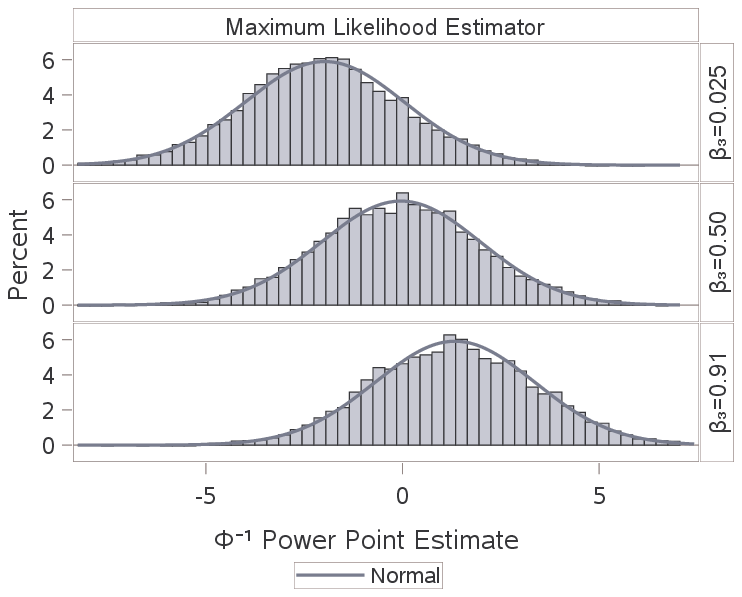}
	\caption{\small{Sampling distribution of the $\Phi^{-1}$ transformed maximum likelihood estimator of power over 10,000 simulations.} }
	\label{sim plot logit}
\end{figure}

\subsection{Extrapolation Between Endpoints or Control Groups Across Phases}\label{extrapolation}
In the examples thus far the phase 2 study used the same endpoint and treatment groups planned for phase 3.  Depending on the therapeutic area and endpoint this may not be feasible.  In such cases 
the phase 3 
treatment effect, and hence phase 3 power, can be transformed into a function of the phase 2 treatment effect.  Of course this modeling brings an additional layer of uncertainty which can be expressed as a confidence band around the power curve.  Figure \ref{cd extrap} shows similar power curves and a confidence density as before, now with a 95\% confidence band around the phase 3 power curve had it been extrapolated from a different phase 3 endpoint or control group.  This extrapolation uncertainty translates into the overall power curve, and easily carries over into Figure \ref{power cd extrap}.  This is a great visual to discern uncertainty around the phase 2 treatment effect and that due to the extrapolation model.  


\begin{figure}[H]
	\centering
	\includegraphics[trim={1cm 18.73cm 0 0}, clip, height = 2.75in]{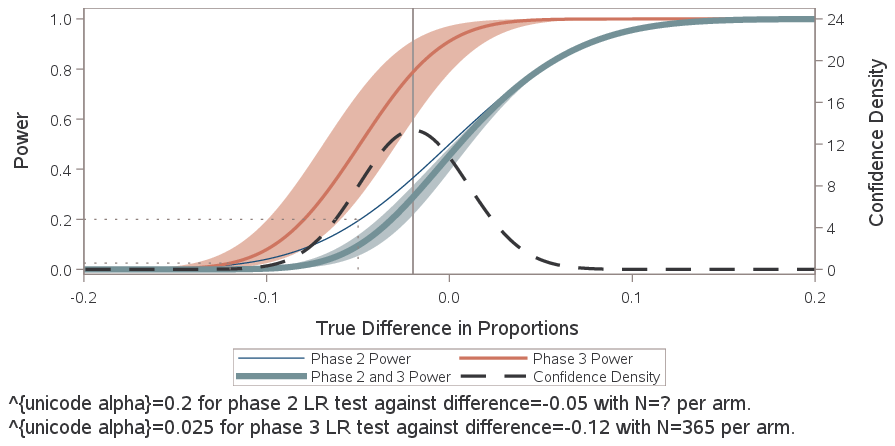}
	\caption{\small{Solid lines depict power curves for a likelihood ratio test of the difference in proportions in phase 2, phase 3, and overall.  Confidence bands depict extrapolation modeling uncertainty.  Dashed line depicts the confidence density for $\theta$ based on historical data and expert opinion.} }
	\label{cd extrap}
\end{figure}

\begin{figure}[H]
	\centering
	\includegraphics[trim={1cm 18cm 0 0}, clip,height =3.05in]{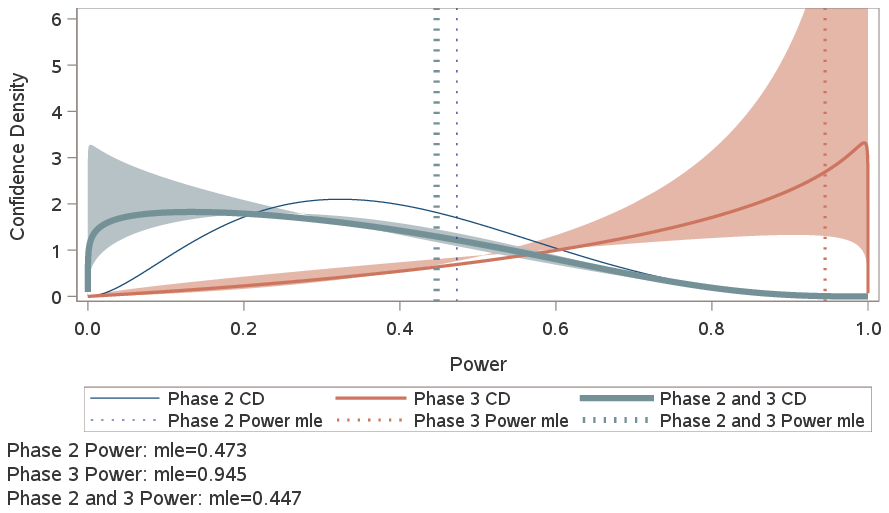} 
	\caption{\small{Solid lines depict resulting confidence densities for power in phase 2, phase 3, and overall.  Dotted lines depict maximum likelihood estimates of power.  Confidence bands depict the extrapolation modeling uncertainty.} }
	\label{power cd extrap}
\end{figure}

For example, suppose the phase 3 study plans to investigate a difference in proportions using a different control therapy than is planned for phase 2.  Suppose further that external studies have been conducted investigating the phase 2 and phase 3 control therapies.  Using a network meta-analysis one can estimate and infer the phase 3 power curve in terms of the phase 2 treatment effect. The population-level treatment effect investigated in phase 2 can be denoted as $\theta_2=p_{active}-p_{ctrl2}$, the population-level difference in proportions between the control therapies can be denoted as $\Delta=p_{ctrl3}-p_{ctrl2}$, and the population-level treatment effect investigated in phase 3 can be denoted as $\theta_3=\theta_2-\Delta=p_{active}-p_{ctrl3}$.  
It is then a simple change of variables to extrapolate the phase 3 power curve $\beta_3(\theta_3)$ in terms of the phase 2 treatment effect, $\beta_3(\theta_2-{\Delta})$.  The function $\beta_3(\cdot)$ is defined by its subscript and not its argument.  Replacing $\Delta$ with a point estimate $\hat{\Delta}$, as well as with lower and upper confidence limits, produces the confidence band around the extrapolated estimated phase 3 power curve, $\beta_3(\theta_2-\hat{\Delta})$.  Similarly, if a p-value function is available for the phase 2 treatment effect, $H(\theta_2)$, replacing $\Delta$ with a point estimate as well as with lower and upper confidence limits produces the confidence band around the p-value function for phase 3 power using the method corresponding to Equation (\ref{power cd}), $H\big(\beta_3^{-1}\{\beta_3(\theta_2-\hat{\Delta})\}+\hat{\Delta}\big)$.  For a given hypothesis for $\theta_2$, the value $H(\theta_2)$ is assigned to $\beta_3(\theta_2-\hat{\Delta})$.  In practice this will be solved numerically in a data step.  To construct a proper p-value function for phase 3 power without confidence bands, one could utilize a transformation of the power point estimate $\beta_3(\hat{\theta}_2-\hat{\Delta})$ along with the delta method and conduct a Wald test.   
To extrapolate between endpoints across phases using external or elicited data that is assumed exchangeable, one could build a regression model of the endpoint planned for phase 3 as a function of the endpoint and treatments planned for phase 2 (or their exchangeable surrogates).  The model contrast statements would then be used to perform a change of variables in the phase 3 power curve similar to that described above.  
Even without extrapolation a similar confidence band visualization can be used to incorporate a confidence interval for a nuisance parameter such as the population-level control therapy response rate when constructing the estimated power curves.  
\\

\section{Adjustment for Multiple Comparisons}\label{multiple comparisons}
Clinical development plans almost always explore multiple endpoints and involve interim analyses, and a natural consideration when discussing frequentist inference is the adjustment for multiple comparisons.  
Even a phase 3 confirmatory setting often involves multiple studies for the explicit purpose of reproducing/replicating results, and regulatory approval can always be changed.  This is to say that if one is capable of updating previously made inference about $\theta$, no adjustment for multiplicity is required.  This perhaps reflects Fisher's position on meta-analysis and inductive reasoning (\citeauthor{lehmann1993} 1993; \citeauthor{efron1998} 1998; \citeauthor{perezgonzalez2015} 2015), 
and is in some ways congruent with objective Bayesianism, though we can not presume to know what Fisher would think if he was alive today.  This viewpoint simply emphasizes the per-comparison error rate knowing no conclusion about $\theta$ is ever final.  Fisher did of course make use of the F-test for what is known in today's terms as controlling a family-wise error rate in the weak sense, and used the entire context of an experiment to determine statistical significance.  P-value functions can certainly be used to display decision rules and study results while adjusting for multiple comparisons if one so chooses to control a particular family-wise error rate.

\section{Additional Figures}\label{additional figures}

\begin{figure}[H]
	\centering
	\includegraphics[trim={1cm 18.7205cm 0 0}, clip, height = 2.75in]{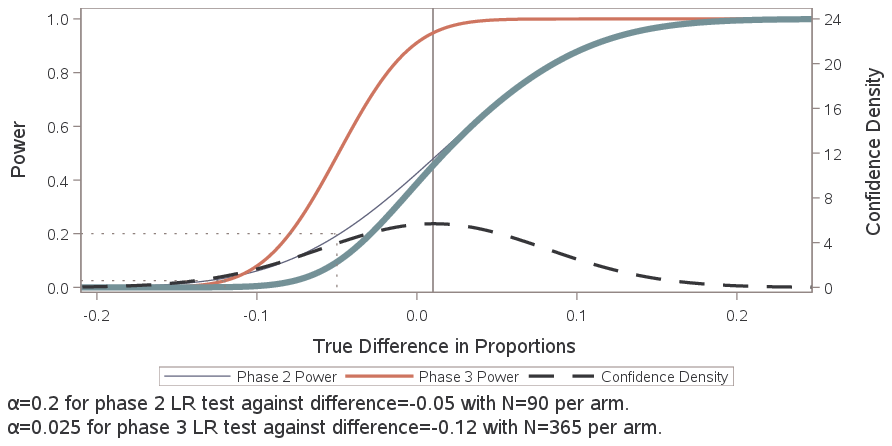}					\caption{\small{Phase 2 power curve testing $H_0$: $\theta\le-0.05$ with N=90 per arm at $\alpha$=0.2.  Phase 3 power curve testing $H_0$: $\theta\le-0.12$ with N=365 per arm at  $\alpha$=0.025.  Confidence density for $\theta$ based on historical data and expert opinion. }} 
\end{figure}

\begin{figure}[H]
	\centering
	\includegraphics[trim={1cm 16.75cm 0 0}, height = 3.3in]{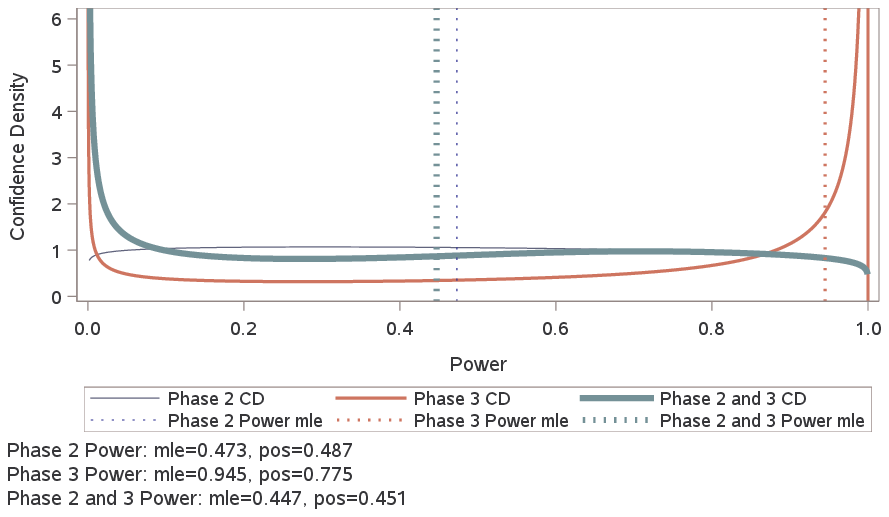} 
	\caption{\small{Solid lines depict resulting confidence distributions for power, $h(\beta)=dH(\theta)/d\beta(\theta)$, in phase 2, phase 3, and overall.  Dotted lines depict maximum likelihood estimates of power.} }
\end{figure}

\begin{figure}[H]
	\centering
	\includegraphics[trim={1cm 18.7205cm 0 0}, clip, height = 2.75in]{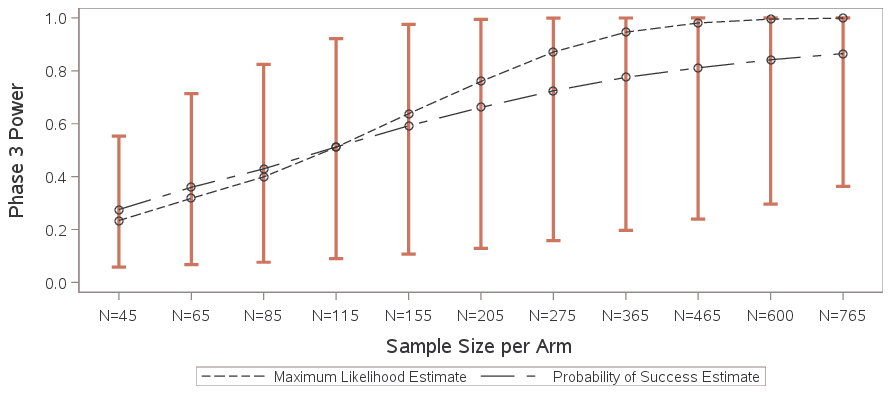}					\caption{\small{Estimated phase 3 power testing $H_0$: $\theta\le-0.12$ at $\alpha$=0.025 at various sample sizes with 80\% confidence limits based on the elicitation (wide). }} 
\end{figure}

\section{Additional Figures}\label{additional figures3}

\begin{figure}[H]
	\centering
\includegraphics[trim={0.8cm 17.5cm 0 0}, clip,height = 3.2in]{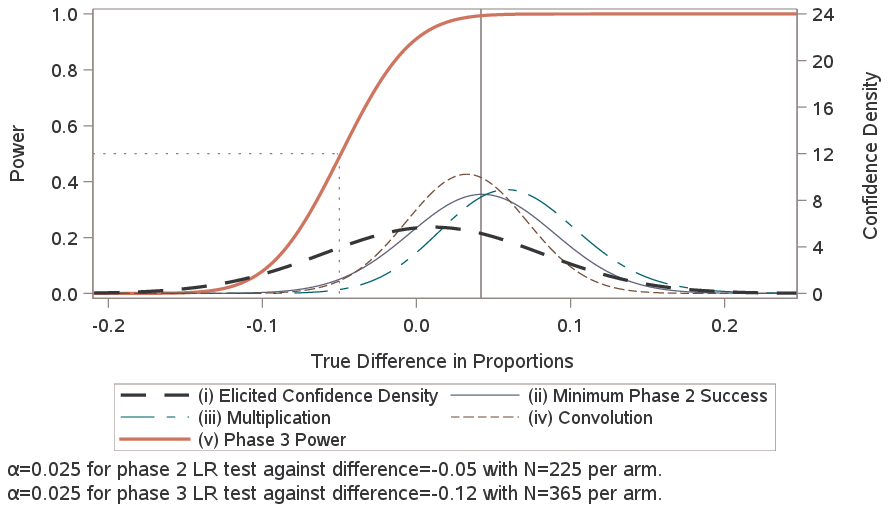} 
	\caption{\small{(i) Elicited confidence density (wide).  (ii) Confidence density for $\theta$ from differentiating the approximate phase 2 power curve  testing $H_0$: $\theta\le-0.05$ with N=225 per arm at $\alpha$=0.025.  (iii) Multiplication of elicited $H(\theta)$ and phase 2 power curve, differentiated.   (iv) Convolution of elicited $H(\theta)$ and approximate phase 2 power curve, differentiated.  (v) Phase 3 power curve testing $H_0$: $\theta\le-0.12$ with N=365 per arm at  $\alpha$=0.025.} }
\end{figure}

\begin{figure}[H]
	\centering
\includegraphics[trim={0.8cm 17.5cm 0 0}, clip,height = 3.2in]{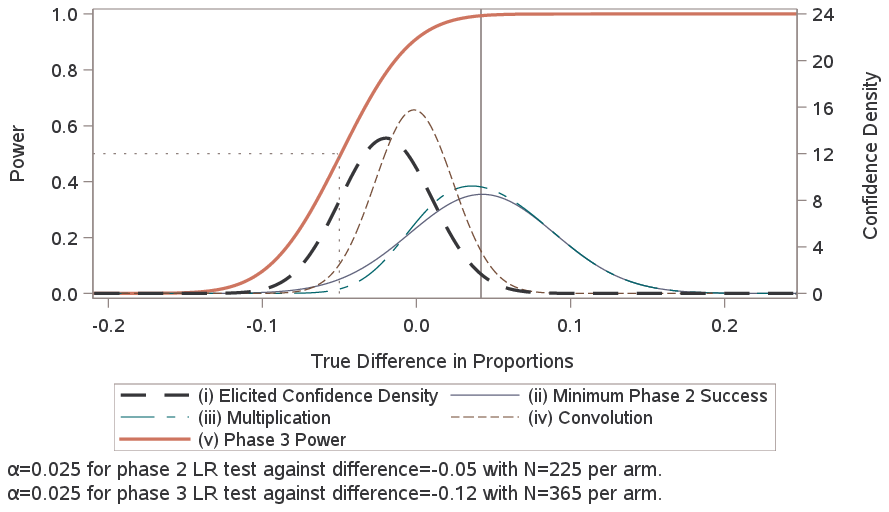} 
	\caption{\small{(i) Elicited confidence density (narrow).  (ii) Confidence density for $\theta$ from differentiating the approximate phase 2 power curve  testing $H_0$: $\theta\le-0.05$ with N=225 per arm at $\alpha$=0.025.  (iii) Multiplication of elicited $H(\theta)$ and phase 2 power curve, differentiated.   (iv) Convolution of elicited $H(\theta)$ and approximate phase 2 power curve, differentiated.  (v) Phase 3 power curve testing $H_0$: $\theta\le-0.12$ with N=365 per arm at  $\alpha$=0.025.} }
\end{figure}

\begin{figure}[H]
	\centering
\includegraphics[trim={0.8cm 17.5cm 0 0}, clip,height = 3.2in]{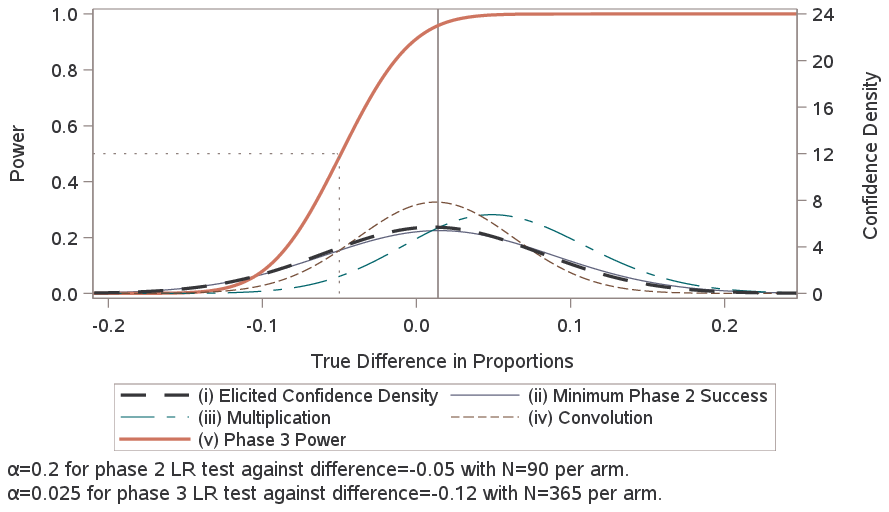} 
	\caption{\small{(i) Elicited confidence density (wide).  (ii) Confidence density for $\theta$ from differentiating the approximate phase 2 power curve  testing $H_0$: $\theta\le-0.05$ with N=90 per arm at $\alpha$=0.2.  (iii) Multiplication of elicited $H(\theta)$ and phase 2 power curve, differentiated.   (iv) Convolution of elicited $H(\theta)$ and approximate phase 2 power curve, differentiated.  (v) Phase 3 power curve testing $H_0$: $\theta\le-0.12$ with N=365 per arm at  $\alpha$=0.025.} }
\end{figure}

\section{{Comparing Distribution Estimates}}\label{construction}

\subsection{Discrete Parameter Space}\label{discrete example}

{When the parameter space is discrete the upper and lower p-value functions $H(\cdot)$ and $H^{-}(\cdot)$ may not form distribution functions on the parameter space.  Nevertheless, these p-value functions are indispensable for performing inference.}  
For example, consider the 3x3 table below depicting the operating characteristics of a cancer screening test with 0.85 specificity and 0.80 sensitivity.  {The parameter space is shown across the top of the table and the support of the sampling distribution (test result) is displayed along the left side of the table so that this table is read vertically.}  If a subject has No Cancer the screening test will produce a Negative result, an At Risk result, and a Positive result $85\%$, $10\%$, and $5\%$ of the time respectively.  Likewise, if the subject indeed has Cancer the test will produce a Negative result, an At Risk result, and a Positive result 5\%, 15\%, and 80\% of the time respectively.  These long-run probabilities can be verified within a margin of error through repeated testing.  {The power of the test shows the ex-ante sampling probability of observing an At Risk or Positive result testing the hypothesis $H_0$: No Cancer as a function of the unknown true cancer status for the subject at hand.  This long-run probability forms the level of confidence in the next observed test result for the subject. } 
\\

The p-value function testing $H_0$: No Cancer, $H_0$: Pre-Cancer, and $H_0$: Cancer as a function of the hypothesis and the observed data is read horizontally and displays the lower-tailed p-value for a Negative result and the upper-tailed p-value for a Positive result.  For an At Risk result the upper-tailed p-value is displayed testing $H_0$: No Cancer and $H_0$: Pre-Cancer, and the lower-tailed p-value is displayed testing $H_0$: Pre-Cancer and $H_0$: Cancer.  If an At Risk result is produced for a given subject, the upper-tailed p-value testing the hypothesis that the subject at hand has No Cancer is the probability of an At Risk or more extreme (Positive) test result given the subject has No Cancer, $0.10 + 0.05=0.15$.  Likewise, for the same At Risk result the lower-tailed p-value testing the hypothesis that the subject at hand has Cancer is the probability of an At Risk or more extreme (Negative) test result given the subject has Cancer, $0.15+0.05=0.20$.  The confidence level is a function of the hypothesis and the observed data.  This table is read horizontally and shows that if the test returns an At Risk result we can ``rule out" $H_0$: No Cancer at the 15\% level and $H_0$: Cancer at the 20\% level and are therefore 65\% confident in the alternative, which is Pre-Cancer.  The 65\% confidence level is nothing more than a restatement of the p-values testing $H_0$: No Cancer and $H_0$: Cancer, $100(1-0.15-0.20)\%$.  Similarly, if the test returns a Positive result we can ``rule out" $H_0$: Pre-Cancer (and by extension $H_0$: No Cancer) at the 10\% level, and are therefore 90\% confident in the alternative, which is Cancer.  Either the subject has Pre-Cancer (or No Cancer) and we have witnessed a 10\% (or smaller) event, or the subject indeed as Cancer.
\\

If a given subject was sampled from an irreducible population that has No Cancer, Pre-Cancer, and Cancer in a 4:2:1 ratio, then the posterior depicts the long-run proportion of cancer status among subjects sampled conditionally from the population given a particular test result.  {In this context these posterior probabilities are often referred to as negative predictive value, false omission rate, false discovery rate, and positive predictive value.}  One way to use this long-run probability to make inference on the cancer status of the subject at hand is by imagining the subject was instead sampled from the posterior distribution.  However, this is a direct contradiction to the earlier claim that the subject at hand was sampled from the prior distribution.  The posterior sampling frame is correct only if the prior sampling frame is correct, yet there can only be a single sampling frame from which we obtained the selected subject at hand.  If the subject really were sampled from a known prior population and no other, this information would be incorporated into the p-value calculation through consideration of the joint distribution. In practice, though, the subject is not sampled from a known prior population and no other.  The Bayesian prior and posterior probabilities might instead be interpreted as measuring the unfalsifiable subjective belief of the experimenter regarding the cancer status of the subject at hand, rather than long-run proportions of cancer status among sampled subjects.  
\\

The likelihood is identified by reading the table of operating characteristics horizontally.  The normalized likelihood can be seen as a posterior based on a 1:1:1 prior.  It is more objectively viewed as an approximate p-value function.  The normalization smooths the operating characteristics of the screening test so the probabilities sum to 1 over the parameter space.  {The plug-in sampling distribution transposes the operating characteristics of the screening test across the parameter space.}  All five methods below use the sampling behavior of the screening test to form a distribution estimate of cancer status.  In this setting the p-values do not form a distribution function on the parameter space.  {If an additional follow-up test is to be conducted on the subject at hand, these distribution estimates can be used to perform inference on the power of the future test.  If one is not satisfied with this inference on power, a more sensitive and specific test can be sought.}  Regardless of paradigm, multiple tests can be performed and the results convolved to improve the inference on the true cancer status for a given subject.    

\begin{table}[H]
	\caption{Cancer Screening Test}
	
	\begin{center}

		\begin{tabular}{ l l c c c }
			&                   &  \multicolumn{3}{c}{True Cancer Status} \\
			 \cline{3-3} \cline{4-4}  \cline{5-5}
			 &  Test Result &  No Cancer  & Pre-Cancer & Cancer   \\
			 &	 &          &      &         \\
			\cline{1-1} \cline{2-2} \cline{3-3} \cline{4-4}  \cline{5-5} 
			 ~ &    ~ &     ~ & ~ &~\\
			 & Negative    &       \multicolumn{1}{c|}{0.85}  &  \multicolumn{1}{|c|}{0.40}	&  \multicolumn{1}{|c}{0.05} \\
			 Operating Characteristics &     At Risk      &     \multicolumn{1}{c|}{0.10}   &	  \multicolumn{1}{|c|}{0.50} &  \multicolumn{1}{|c}{0.15}	\\
			 & Positive &    \multicolumn{1}{c|}{0.05}   &  \multicolumn{1}{|c|}{0.10} &	 \multicolumn{1}{|c}{0.80} \\

			  &     &       &  &  \\
			  &     &       &  &  \\

			Power  & &0.15    &0.60   &0.95 \\

			  &    &       &  &  \\
			  &     &       &  &  \\

			  & Negative    &       0.85 &0.40	& 0.05 \\
			\cline{3-3} \cline{4-4}  \cline{5-5} 
			One-sided p-value &At Risk& 0.15    & 0.60$|$0.90	 &0.20	\\
			\cline{3-3} \cline{4-4}  \cline{5-5} 
			 (Confidence Curve)& Positive &   0.05 &0.10	& 0.80 \\

			  &     &       &  &  \\
			  &     &       &  &  \\

			  & Negative    &       \textit{0.60} &0.40	& 0.05 \\
			\cline{3-3} \cline{4-4}  \cline{5-5} 
			 Confidence Level &At Risk& 0.15    & \textit{0.65}	 &0.20	\\
			\cline{3-3} \cline{4-4}  \cline{5-5} 
			 & Positive & 0.05   &0.10	& \textit{0.90} \\

			  &     &       &  &  \\
			  &     &       &  &  \\

			 & Negative    &       0.80	& 0.19 & 0.01 \\
			\cline{3-3} \cline{4-4}  \cline{5-5} 
			 Posterior (4:2:1 Prior) &   At Risk         & 0.26    &0.65	 &0.10	\\
			\cline{3-3} \cline{4-4}  \cline{5-5} 
			 & Positive &   0.17 &0.17	& 0.67 \\

			  &     &       &  &  \\
			  &     &       &  &  \\

			 & Negative    &       0.65	&0.31 & 0.04 \\
			\cline{3-3} \cline{4-4}  \cline{5-5} 
			 Normalized Likelihood &     At Risk       &0.13     &0.67	 &0.20	\\
			\cline{3-3} \cline{4-4}  \cline{5-5} 
			 & Positive &   0.05 &0.11	& 0.84 \\
			  &     &       &  &  \\
			  &     &       &  &  \\

			& Negative    &       0.85	&0.10 & 0.05 \\
			\cline{3-3} \cline{4-4}  \cline{5-5} 
			 Plug-in Sampling Distribution &     At Risk       &0.40     &0.50	 &0.10	\\
			\cline{3-3} \cline{4-4}  \cline{5-5} 
			 & Positive &   0.05 &0.15	& 0.80 \\

			 & &  & &\\
			\cline{1-1} \cline{2-2} \cline{3-3} \cline{4-4}  \cline{5-5}
		\end{tabular}
	\end{center}
	\small{\space\space\space }
\end{table}

\subsection{Distribution Estimates Giving Different Results}
\begin{figure}[H]
	\centering
		{\includegraphics[trim={0.75cm 17.5cm 0 0},clip, height =3.2in]{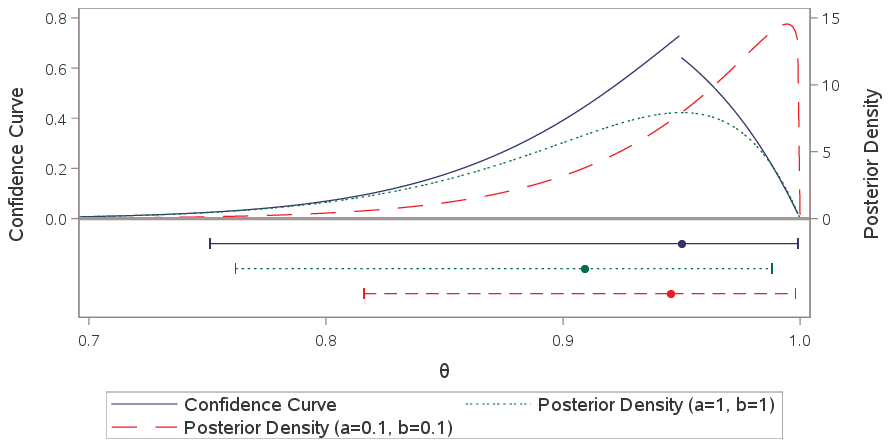}}
	\caption{\small{Exact frequentist and Bayesian inference on a binomial proportion $\theta$ based on a sample of size $n=20$.} }
	\label{bayes vs freq}
\end{figure}

{Let $X_1,...,X_n\sim \text{Bernoulli}(\theta)$.  The confidence curve and 95\% confidence interval in Figure} \ref{bayes vs freq} {show exact inference on $\theta$ from inverting the cumulative distribution function for $\sum{X}\sim \text{Bin}(n,\theta)$ based on a sample of size $n=20$ with $\sum x=19$ events.  In this setting the conjugate Bayesian prior is a $\text{Beta}(a,b)$ distribution.  The green dotted density shows a Bayesian posterior and 95\% credible interval based on a non-informative $\text{Beta}(1,1)$ prior.  The red dashed density shows a Bayesian posterior and 95\% credible interval based on a non-informative $\text{Beta}(0.1,0.1)$ prior.  The $\text{Beta}(0.1,0.1)$ prior has a larger variance compared to the uniform prior yet it produces shorter posterior credible intervals.  While the uniform prior produces the widest possible objective posterior intervals, they are noticeably shorter than the corresponding exact confidence intervals.  Additionally, the posterior mean as an estimator for $\theta$ based on a uniform prior is biased towards 0.5.  With increasing sample size all three distribution estimators will produce similar results.}

\subsection{From Confidence Intervals to Distribution Estimates}\label{construction of densities}

	
\begin{figure}[H]
	\centering
	{\includegraphics[trim={0cm 15.5cm 0 0}, height = 3.3in]{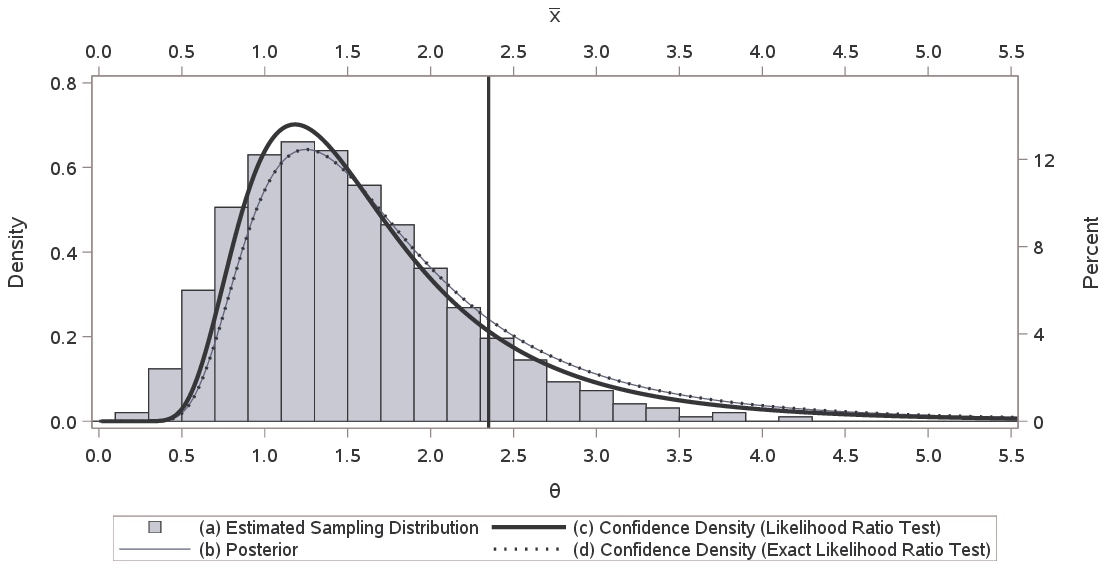}}
	\caption{\small{(a) Plug-in estimated sampling distribution for the MLE of the mean supported by 
$\bar{x}$ for exponentially distributed data with $n=5$, replacing the unknown fixed true $\theta$ with $\hat{\theta}_{mle}$=1.5.  (b) Bayesian posterior from vague conjugate prior supported by $\theta$.  (c) Confidence distribution (density) based on the likelihood ratio test supported by $\theta$.  (d) Confidence distribution (density) based on the exact likelihood ratio test supported by $\theta$.} }
	\label{priors}
\end{figure}
Consider the setting where $X_1,...,X_n\sim\text{Exp}(\theta)$ with likelihood function $L(\theta)=\theta^{-n} e^{-\sum{x_i}/\theta}$.  
Then $supL(\theta)$ yields $\hat{\theta}_{mle}=\bar{x}$ as the maximum likelihood estimate for $\theta$, the likelihood ratio test statistic is $-2\text{log}LR({\boldsymbol{x},\theta_0})\equiv-2\text{log}\big(L({\theta}_0)/L(\hat{\theta}_{mle}) \big)$, and the corresponding upper p-value function (and confidence distribution function) is defined as in Equation (\ref{eq}).  The histogram in Figure \ref{priors}, supported by 
$\bar{x}$, depicts the plug-in estimated sampling distribution for the maximum likelihood estimator (MLE) of the mean for exponentially distributed data with $n=5$ based on $\hat{\theta}_{mle}=1.5$.  
Replacing the unknown fixed true $\theta$ with $\hat{\theta}_{mle}=1.5$, this displays the estimated sampling behavior of the MLE for all other replicated experiments, a $\text{Gamma}(n,\hat{\theta}_{mle}/n)$ distribution.  The Bayesian posterior 
depicted by the thin blue curve resulting from a vague conjugate prior or an improper $1/\theta$ prior is a transformation of the likelihood 
and is supported on the parameter space, an Inverse Gamma(5,7.5) distribution. The bold black curve is also data dependent and supported on the parameter space, but represents confidence intervals of all levels from inverting the likelihood ratio test.  It is a transformation of the sampling behavior of the test statistic under the null onto the parameter  space, a ``distribution" of p-values.  Each value in the parameter space takes its turn playing the role of null hypothesis and hypothesis testing (akin to proof by contradiction) is used to infer the unknown fixed true $\theta$.  The area under this curve to the right of the reference line is the p-value or significance level when testing the hypothesis $H_0$: $\theta \ge 2.35$.  This probability forms the level of confidence that $\theta$ is greater than or equal to $2.35$.  Similarly, the area to the left of the reference line is the p-value when testing the hypothesis $H_0$: $\theta \le 2.35$.  
One can also identify the two-sided equal-tailed $100(1-\alpha)\%$ confidence interval by finding the complement of those values of $\theta$ in each tail with $\alpha$/2 significance.   
The dotted curve shows the exact likelihood ratio confidence density formed by noting that $\bar{X}\sim$ Gamma$(n,\theta/n)$ and inverting its cumulative distribution function.  This confidence density coincides perfectly with the posterior distribution.  A confidence density similar to that based on the likelihood ratio test can be produced by utilizing a Wald test with a log link.

\begin{figure}[H]
	\centering
	{\includegraphics[trim={1.2cm 18.85cm 0 0},clip, height =2.6in]{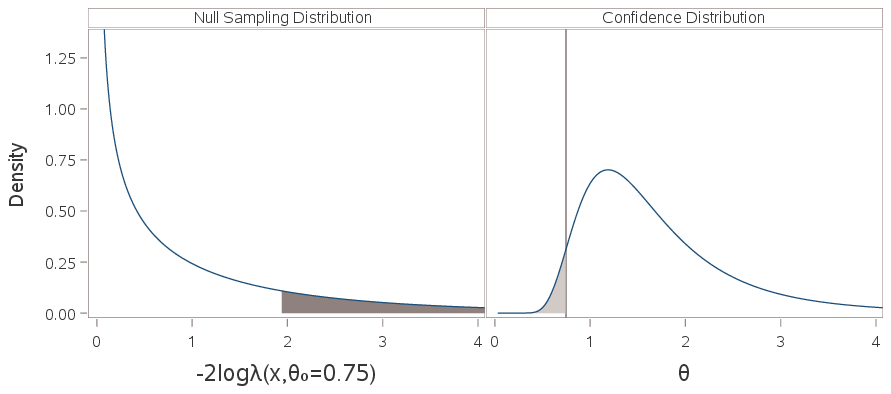}}
	\caption{\small{Approximate $\chi^2_1$  null sampling distribution of the likelihood ratio test statistic for testing $H_0$: $\theta \le 0.75$.} }
\label{chi_approx}
\end{figure}

\begin{figure}[H]
	\centering
	{\includegraphics[trim={1.2cm 18.85cm 0 0},clip, height =2.6in]{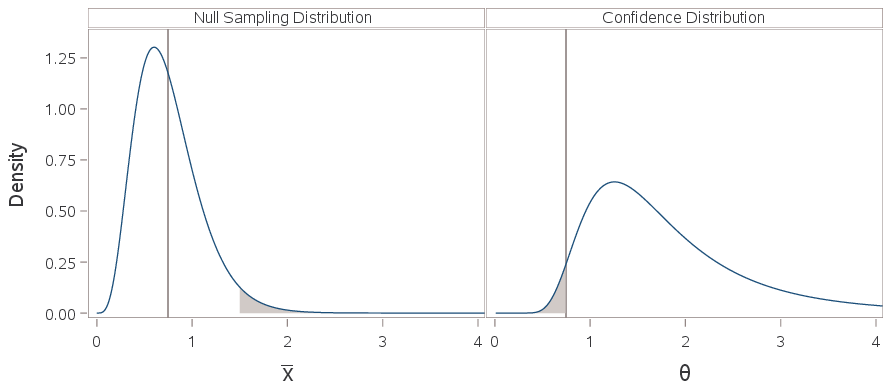}}
	\caption{\small{Exact null sampling distribution of $\hat{\theta}_{MLE}=\bar{X}$ for testing $H_0$: $\theta \le 0.75$.} }
\label{gamma_exact}
\end{figure}
$H(\theta)$ captures the upper-tailed  p-value for every value of $\theta$ in the parameter space, and $dH(\theta)/d\theta$  is the resulting confidence density.  The confidence density in Figure \ref{priors} was constructed using the $\chi^2_1$ approximation for the sampling distribution of the likelihood ratio test statistic.  In Figure \ref{chi_approx} the 2-sided p-value testing $H_0$: $\theta =0.75$ is shaded in the left panel.  Half of this is the one-sided p-value testing $H_0$: $\theta \le 0.75$.  This is shaded above $\theta \le 0.75$ in the right panel.  A single $\chi^2_1$ reference distribution is used, and the value of the test statistic depends on the hypothesis being tested.  This approximation is particularly useful when considering differences in parameters or other more complicated functions.  When performing inference on an exponential rate parameter one can note the likelihood ratio test statistic is a monotonic function of $\hat{\theta}_{MLE}=\bar{X}$, which follows a Gamma($n$, $\theta$/$n$) distribution.   Referencing this distribution allows the calculation of the exact likelihood ratio test p-value.  In Figure \ref{gamma_exact} the left panel shows the null sampling distribution when testing $H_0$: $\theta \le 0.75$.  The one-sided p-value in the left panel is shaded above $\theta \le 0.75$ in the right panel.  The location of the null sampling distribution depends on the hypothesis being tested.

\subsection{Distribution Estimates for Meta-Analysis}


\begin{figure}[H]
	\centering
	{\includegraphics[trim={1.2cm 18cm 0 0}, height = 3in]{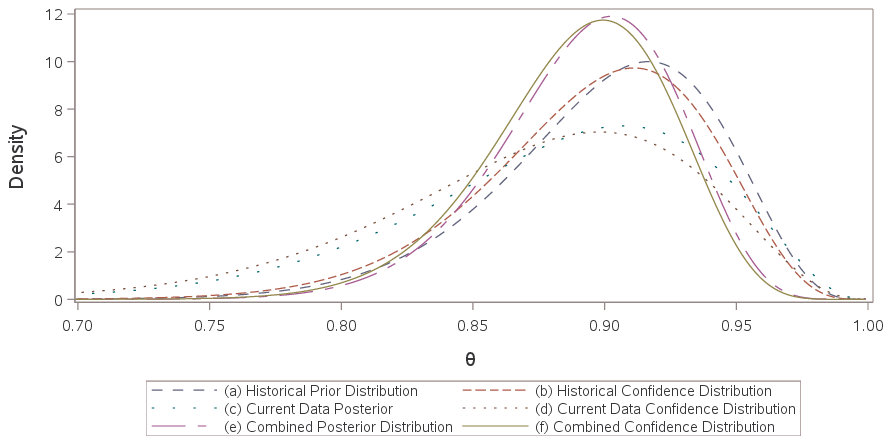}} 
	\caption{\small{(a) Informative Bayesian prior distribution based on historical likelihood and vague conjugate prior for binomial proportion, $\hat{\theta}_{Bayes}^{Hist}=0.90, n=50$.  (b) Confidence distribution (likelihood ratio test) based on historical data for binomial proportion, $\hat{\theta}_{mle}^{Hist}=0.90, n=50$.  (c) Bayesian posterior based on current likelihood and vague conjugate prior, $\hat{\theta}_{Bayes}^{Current}=0.87, n=30$.  (d) Confidence distribution (likelihood ratio test) based on current data, $\hat{\theta}_{mle}^{Current}=0.87, n=30$.  (e) Posterior distribution based on informative historical prior and current data likelihood.  (f) Convolution of historical and current confidence distributions.} }
	\label{confidence distributions}
\end{figure}

Figure 2 depicts a meta-analysis using confidence distributions for a binomial proportion $\theta$.  Density (a) represents an informative Bayesian prior distribution based on a historical likelihood and a vague conjugate prior producing an estimate of 0.90 from a sample size of $n=50$.  This same information is depicted in (b) as a confidence density resulting from a likelihood ratio test.  A similar confidence density can be produced by utilizing a Wald test with a logit link.  The Bayesian posterior based on the current data binomial likelihood and a vague conjugate prior is shown in (c) with an estimate of 0.87 resulting from $n=30$.  This same information can be represented as a likelihood ratio confidence density, (d).  Using Bayes' theorem, the prior (a) and the likelihood from (c) combine to form (e).  Using the convolution formula in Equation (\ref{conf_dist_2}), (b) and (d) combine to form (f).      

\subsection{Bayesian versus Frequentist Interpretations of Probability}
In any quantitative field it is not enough to simply apply a set of mathematical operations.  One must also provide an interpretation.  The field of statistics concerns itself with a special branch of mathematics regarding probability.  When interpreting probability there are primarily two competing paradigms: frequentist and Bayesian.  These paradigms differ on what it means for something to be considered random and what probability itself measures.  Both frequentists and Bayesians would agree that once a test statistic is observed it is fixed, there is nothing random about it.  Additionally, frequentists and most Bayesians would agree that the $\theta$ under investigation is an unknown fixed quantity and it is simply treated as random in the Bayesian paradigm as a matter of practice.  The question then becomes, ``How do we interpret probability statements about a fixed quantity?"  Without delving into the mathematical details of how a posterior or a p-value is calculated, we explore various interpretations below and what makes them untenable. 
\\

One interpretation of a Bayesian prior is that ``random'' is synonymous with ``unknown'' and probability measures the experimenter's belief (``knowledge," ``judgment," ``opinion," etc.) so that the posterior measures belief about the unknown fixed true $\theta$ given the observed data.  This interpretation is untenable because belief is unfalsifiable $-$ it is not an empirically investigable statement about the parameter, a hypothesis, nor an experiment.  Another interpretation is that ``random'' is short for ``random variable'' (the act of sampling mapped to the real line) and probability measures the emergent pattern of many samples so that a Bayesian prior is merely a modeling assumption regarding $\theta$, i.e. the unknown fixed true $\theta$ was sampled from a known collection or prevalence of $\theta$'s (prior distribution) and the observed data is used to subset this collection, forming the posterior distribution (\citeauthor{good1965} 1965, \nocite{good1966}1966).  
The unknown fixed true $\theta$ is now imagined to have instead been sampled from the posterior.  This interpretation is untenable because of the contradiction caused by claiming two sampling frames.  The second sampling frame is correct only if the first sampling frame is correct, yet there can only be a single sampling frame from which we obtained the unknown fixed true $\theta$ under investigation.  A third interpretation of a Bayesian prior is that ``random'' is synonymous with ``unrealized'' or ``undetermined'' and probability measures a simultaneity of existance so that $\theta$ is \textit{not} fixed and all values of $\theta$ are true simultaneously; the truth exists in a superposition depending on the data observed according to the posterior distribution (\citeauthor{schrodinger1980} 1980; \citeauthor{ballentine1970} 1970).   This interpretation is untenable because it reverses cause and effect $-$ the population-level parameter depends on the data observed, but the observed data depended on the parameter.  Ascribing any of these interpretations to the posterior allows one to make philosophical probability statements about hypotheses given the data.  While the p-value is typically not interpreted in the same manner, it does show us the plausibility of a hypothesis given the data $-$ the ex-post sampling probability of the observed result or something more extreme if the hypothesis for the unknown fixed $\theta$ is true.  When interpreting a small p-value, either the null hypothesis is true and we have witnessed a rare event, or the null hypothesis is false.  These statements are empirically investigable within a margin of error through repeated sampling.    
\\

One might notice the similarity between a confidence distribution (or more generally a p-value function) and a posterior distribution and wonder under what circumstances is each one preferable.  At its essence this is a matter of scientific objectivity (\citeauthor{efron1986} 1986).  To the Bayesian, probability is axiomatic and measures the experimenter.  To the frequentist, probability measures the experiment and must be investigable.  The Bayesian interpretation of probability as a measure of belief is unfalsifiable.  Only if there exists a real-life mechanism by which we can sample values of $\theta$ can a probability distribution for $\theta$ be investigated.  In such settings, probability statements about $\theta$ would have a purely frequentist interpretation.  This may be a reason why frequentist inference is ubiquitous in the scientific literature.  If the prior distribution is chosen in such a way that the posterior is dominated by the likelihood or is proportional to the likelihood, Bayesian belief is more objectively viewed as a crude p-value function.  
\\

In short, for those who subscribe to the frequentist interpretation of probability, the p-value function summarizes all the probability statements about the experiment one can make.  
It is a matter of correct interpretation given the definition of probability and what constitutes a random variable.  The posterior remains an incredibly useful tool and can be interpreted as an approximate p-value function.

\section{SAS Code}\label{sas code}
\footnotesize
\begin{lstlisting}
%let ddiff=0.001;

data binomial;

*diff is the theta axis, the true difference in proportions;
do diff=-0.21 to 0.247 by &ddiff.;

*Elicitation;

int_ctrl=0.43;
diff_hat=-0.02;
n_ctrl=1200;
n_active=350;

y_ctrl=int_ctrl*n_ctrl;
y_active=(int_ctrl+diff_hat)*n_active;

*Wald CD;
/*
p_active=y_active/n_active;
se=sqrt( p_active*(1-p_active)/n_active + int_ctrl*(1-int_ctrl)/n_ctrl );
H=1-cdf('normal',(diff_hat-diff)/(se),0,1);
*/

*Likelihood Ratio Test;

int_ctrl_null=(y_ctrl+(y_active/(int_ctrl+diff))*int_ctrl-((y_active/(int_ctrl+diff))
	*int_ctrl)*int_ctrl+(y_active*(1-int_ctrl)/(1-int_ctrl-diff))*int_ctrl)
	/(n_ctrl+(n_active*(1-int_ctrl)/(1-int_ctrl-diff)));

do i=1 to 100;
int_ctrl_null=(y_ctrl+(y_active/(int_ctrl_null+diff))*int_ctrl_null-((y_active
	/(int_ctrl_null+diff))*int_ctrl_null)*int_ctrl_null+(y_active*(1-int_ctrl_null)
	/(1-int_ctrl_null-diff))*int_ctrl_null)/(n_ctrl+(n_active*(1-int_ctrl_null)
	/(1-int_ctrl_null-diff)));
end;


lambda=((int_ctrl_null/int_ctrl)**y_ctrl)*(( (1-int_ctrl_null)/ (1-int_ctrl)  )
	**(n_ctrl-y_ctrl))*(( (int_ctrl_null+diff)/(int_ctrl+diff_hat)  )**y_active)
	*(((1-int_ctrl_null-diff)/ (1-int_ctrl-diff_hat) )**(n_active-y_active));
loglambda=log(lambda);
minus2loglambda=-2*loglambda;

if diff gt diff_hat then do;
H=(1+cdf('chisquare',-2*loglambda,1))/2;
end;

if diff le diff_hat then do;
H=(1-cdf('chisquare',-2*loglambda,1))/2;
end;


dHddiff=(H-lag(H))/(diff-lag(diff));

C=H*(diff lt diff_hat) + (1-H)*(diff gt diff_hat);

*Phase 2;

n_active_phase2=90; call symput('n_active_phase2',trim(left(n_active_phase2)));

n_ctrl_phase2=90; call symput('n_ctrl_phase2',trim(left(n_ctrl_phase2)));


*null hypothesis;
lower_margin2=-0.05; call symput('lower_margin_phase2',strip(lower_margin2));
alpha_phase2=0.20; call symput('alpha_phase2',strip(alpha_phase2));
*critical effect size; 
lower_cv2=lower_margin2+0.064; call symput('phase2_mle_success',lower_cv2);


y_ctrl_phase2=int_ctrl*(n_ctrl_phase2);
y_active_phase2=(int_ctrl+lower_cv2)*n_active_phase2;

*Wald;
/*p_active_phase2=y_active_phase2/n_active_phase2;
p_ctrl_phase2=y_ctrl_phase2/n_ctrl_phase2;
se_phase2=sqrt( p_active_phase2*(1-p_active_phase2)/n_active_phase2 + p_ctrl_phase2
	*(1-p_ctrl_phase2)/n_ctrl_phase2 );
phase2_power=1-cdf('normal',(p_active_phase2-p_ctrl_phase2-diff)/se_phase2,0,1);*/


*Likelihood Ratio Test;

int_ctrl_null=(y_ctrl_phase2+(y_active_phase2/(int_ctrl+diff))*int_ctrl
	-((y_active_phase2/(int_ctrl+diff))*int_ctrl)*int_ctrl+(y_active_phase2
	*(1-int_ctrl)/(1-int_ctrl-diff))*int_ctrl)/(n_ctrl_phase2+(n_active_phase2
	*(1-int_ctrl)/(1-int_ctrl-diff)));

do i=1 to 100;
int_ctrl_null=(y_ctrl_phase2+(y_active_phase2/(int_ctrl_null+diff))*int_ctrl_null
	-((y_active_phase2/(int_ctrl_null+diff))*int_ctrl_null)*int_ctrl_null
	+(y_active_phase2*(1-int_ctrl_null)/(1-int_ctrl_null-diff))*int_ctrl_null)
	/(n_ctrl_phase2+(n_active_phase2*(1-int_ctrl_null)/(1-int_ctrl_null-diff)));
end;

likelihood_phase2=(int_ctrl_null**y_ctrl_phase2)*(1-int_ctrl_null)**(n_ctrl_phase2
	-y_ctrl_phase2)*((int_ctrl_null+diff)**(y_active_phase2))
	*((1-int_ctrl_null-diff)**(n_active_phase2-y_active_phase2));
likelihood_1_phase2=(int_ctrl**y_ctrl_phase2)*(1-int_ctrl)**(n_ctrl_phase2
	-y_ctrl_phase2)*((int_ctrl+lower_cv2)**(y_active_phase2))*((1-int_ctrl
	-lower_cv2)**(n_active_phase2-y_active_phase2));

lambda_phase2=( likelihood_phase2 )/( likelihood_1_phase2   );
loglambda_phase2=log(lambda_phase2);
minus2loglambda_phase2=-2*loglambda_phase2;

if diff lt lower_cv2 then do;
	phase2_power=(1-cdf('chisquare',-2*loglambda_phase2,1))/2;
end;
else if diff ge lower_cv2 then do;
	phase2_power=(1+cdf('chisquare',-2*loglambda_phase2,1))/2;
end;



*CD for definition of success;
H_phase2_success=phase2_power;

dH_phase2_success_ddiff=(H_phase2_success-lag(H_phase2_success))/(diff-lag(diff));

C_phase2_success=H_phase2_success*(diff lt lower_cv2) + (1-H_phase2_success)*(diff gt lower_cv2);


*Phase 3;

n_ctrl_phase3=365; call symput('n_ctrl_phase3',trim(left(n_ctrl_phase3)));
n_active_phase3=365; call symput('n_active_phase3',trim(left(n_active_phase3)));

*null hypothesis;
lower_margin3=-0.12;  call symput('lower_margin_phase3',strip(lower_margin3));
alpha_phase3=0.025; call symput('alpha_phase3',strip(alpha_phase3));
*minimum detectable effect;
lower_cv3=lower_margin3+0.071;


y_ctrl_phase3=int_ctrl*n_ctrl_phase3;
y_active_phase3=(int_ctrl+lower_cv3)*n_active_phase3;

*Wald;
/*p_active_phase3=y_active_phase3/n_active_phase3;
p_ctrl_phase3=y_ctrl_phase3/n_ctrl_phase3;
se_phase3=sqrt( p_active_phase3*(1-p_active_phase3)/n_active_phase3 
	+ p_ctrl_phase3*(1-p_ctrl_phase3)/n_ctrl_phase3 );
phase3_power1=1-cdf('normal',(p_active_phase3-p_ctrl_phase3-diff)/se_phase3,0,1);*/


*Likelihood Ratio Test;

int_ctrl_null=(y_ctrl_phase3+(y_active_phase3/(int_ctrl+diff))*int_ctrl
	-((y_active_phase3/(int_ctrl+diff))*int_ctrl)*int_ctrl+(y_active_phase3
	*(1-int_ctrl)/(1-int_ctrl-diff))*int_ctrl)/(n_ctrl_phase3+(n_active_phase3
	*(1-int_ctrl)/(1-int_ctrl-diff)));

do i=1 to 100;
int_ctrl_null=(y_ctrl_phase3+(y_active_phase3/(int_ctrl_null+diff))*int_ctrl_null
	-((y_active_phase3/(int_ctrl_null+diff))*int_ctrl_null)*int_ctrl_null
	+(y_active_phase3*(1-int_ctrl_null)/(1-int_ctrl_null-diff))*int_ctrl_null)
	/(n_ctrl_phase3+(n_active_phase3*(1-int_ctrl_null)/(1-int_ctrl_null-diff)));
end;

likelihood_phase3=(int_ctrl_null**y_ctrl_phase3)*(1-int_ctrl_null)**(n_ctrl_phase3
	-y_ctrl_phase3)*((int_ctrl_null+diff)**(y_active_phase3))*((1-int_ctrl_null
	-diff)**(n_active_phase3-y_active_phase3));
likelihood_1_phase3=(int_ctrl**y_ctrl_phase3)*(1-int_ctrl)**(n_ctrl_phase3
	-y_ctrl_phase3)*((int_ctrl+lower_cv3)**(y_active_phase3))*((1-int_ctrl
	-lower_cv3)**(n_active_phase3-y_active_phase3));
lambda_phase3=( likelihood_phase3 )/( likelihood_1_phase3   );
loglambda_phase3=log(lambda_phase3);
minus2loglambda_phase3=-2*loglambda_phase3;

if diff lt lower_cv3 then do;
	phase3_power=(1-cdf('chisquare',-2*loglambda_phase3,1))/2;
end;
else if diff ge lower_cv3 then do;
	phase3_power=(1+cdf('chisquare',-2*loglambda_phase3,1))/2;
end;
if phase3_power=0 then phase3_power=.;


*CD for definition of success;
H_phase3_success=phase3_power;
dH_phase3_success_ddiff=(H_phase3_success-lag(H_phase3_success))/(diff-lag(diff));

C_phase3_success=(H_phase3_success)*(diff lt lower_cv3) + (1-H_phase3_success)*(diff gt lower_cv3);


*CDs for Power.  Derivative of H wrt power;

phase2and3_power=phase3_power*phase2_power;


dH_dpower=(H-lag(H))/(phase3_power-lag(phase3_power));
	if dH_dpower =0 then dH_dpower=.;
	if 0 gt phase3_power gt 1 then dH_dpower=.;

	

dH_dphase2power=(H-lag(H))/(phase2_power-lag(phase2_power));	
	if dH_dphase2power =0 then dH_dphase2power=.;
	if 0 gt phase2_power gt 1 then dH_dphase2power=.;

	
	
dH_dphase23power=(H-lag(H))/(phase2and3_power-lag(phase2and3_power));	
	if 0 gt phase2and3_power gt 1 then dH_dphase23power=.;
	

dH_phase2_dphase3=(H_phase2_success-lag(H_phase2_success))/(phase3_power
	-lag(phase3_power));	

	

*Additional phase 2 inference;	

H_multiply=H*H_phase2_success;
dH_multiply_ddiff=(H_multiply-lag(H_multiply))/(diff-lag(diff));
C_multiply=H_multiply*(H_multiply lt 0.5) + (1-H_multiply)*(H_multiply gt 0.5);


elicited_var=(diff_hat+int_ctrl)*(1-diff_hat-int_ctrl)/n_active+(int_ctrl)
	*(1-int_ctrl)/n_ctrl;
phase2_var=(lower_cv2+int_ctrl)*(1-lower_cv2-int_ctrl)/n_active_phase2
	+int_ctrl*(1-int_ctrl)/n_ctrl_phase2;
phase3_var=(lower_cv3+int_ctrl)*(1-lower_cv3-int_ctrl)/n_active_phase3
	+int_ctrl*(1-int_ctrl)/n_ctrl_phase3;
H_convolve=cdf('normal',(quantile('normal',H,0,1)/sqrt( elicited_var )
	+quantile('normal',H_phase2_success,0,1)/sqrt(  phase2_var))
	/sqrt(1/( elicited_var ) + 1/(phase2_var )  ),0,1);
dH_convolve_ddiff=(H_convolve-lag(H_convolve))/(diff-lag(diff));
C_convolve=H_convolve*(H_convolve lt 0.5) + (1-H_convolve)*(H_convolve gt 0.5);

*Weights for PoS calculations;	
weight=H-lag(H);

weight_phase3cond2=H_phase2_success-lag(H_phase2_success);


*Reference lines and shaded regions in figures;	
if diff le lower_margin2 then ref1=alpha_phase2; else ref1=.;
if diff le lower_margin3 then ref2=alpha_phase3; else ref2=.;
if phase2_power le alpha_phase2 then ref3=lower_margin2;	else ref3=.;
if phase3_power le alpha_phase3 then ref4=lower_margin3; else ref4=.;

if diff le lower_margin2 then ref5=0.5; else ref5=.;
if phase3_power le 0.5 then ref6=lower_margin2; else ref6=.;



if 0.49 le phase2_power le 0.51 then do; call symput('tail2',H); end;
if 0.49 le phase3_power le 0.51 then do; call symput('tail3',H); end;
if 0.49 le phase2and3_power le 0.51 then do; call symput('tail23',H); end;

if H_phase2_success le alpha_phase2 then area=dH_phase2_success_ddiff; 
else area=.;
if H_phase2_success le alpha_phase2 then area2=dH_phase2_dphase3; 
else area2=.;
	

output;
end;
run;


*PoS Calculations;
proc means data=binomial mean noprint;
weight weight;
var phase2_power phase3_power phase2and3_power;
output out=mean_power (where=(_stat_='MEAN'));
run;

proc means data=binomial  mean noprint;
weight weight_phase3cond2;
var phase3_power ;
output out=mean_phase3cond2_power mean=phase3cond2_power;
run;

data mean_power;
set mean_power;
call symput('mean_phase2_power',strip(round(phase2_power,0.001)));
call symput('mean_phase3_power',strip(round(phase3_power,0.001)));
call symput('mean_phase23_power',strip(round(phase2and3_power,0.001)));
run;

data mean_phase3cond2_power;
set mean_phase3cond2_power;
call symput('mean_phase3cond2_power',strip(round(phase3cond2_power,0.001)));
run;






*MLEs;
proc sql noprint;
select diff_hat
into: diff_hat
from binomial;
quit;

proc means data=binomial    noprint;
where &diff_hat.-&ddiff. le diff le &diff_hat.+&ddiff.;
var phase2_power phase3_power phase2and3_power;
output out=mles_power (where=(_stat_='MIN'));
run;

proc means data=binomial    noprint;
where 0.5-0.01 le H_multiply le 0.5+0.01;
var  phase3_power ;
output out=mle_phase3cond2_power (where=(_stat_='MIN'));
run;

data mles_power;
set mles_power;
call symput('phase2_power_mle',strip(round(phase2_power,0.001)));
call symput('phase3_power_mle',strip(round(phase3_power,0.001)));
call symput('phase23_power_mle',strip(round(phase2and3_power,0.001)));
run;

data mle_phase3cond2_power;
set mle_phase3cond2_power;
call symput('phase3cond2_power_mle',strip(round(phase3_power,0.001)));
run;





*Prepare for plots;
data binomial_stack;
set binomial (in=a) binomial (in=b);
if a then do;
	phase=2;
	dH_success_ddiff=dH_phase2_success_ddiff;
	C_success=C_phase2_success;
	mle=lower_cv2;
	lower_margin=lower_margin2;
end;
if b then do;
	phase=3;
	dH_success_ddiff=dH_phase3_success_ddiff;
	C_success=C_phase3_success;
	mle=lower_cv3;
	lower_margin=lower_margin3;
end;
run;

*Check type I error rates;	

proc sql noprint;
select max(phase2_power)
into: cp_phase2
from binomial_stack
where lower_margin2-&ddiff. < diff < lower_margin2 + &ddiff. and phase=2;

select max(phase3_power)
into: cp_phase3
from binomial_stack
where lower_margin3-&ddiff. < diff < lower_margin3 + &ddiff. and phase=3;

quit;
%put &cp_phase2. &cp_phase3.;


*Plots;

proc format;
value phase 2='Phase 2 Success'
	    3='Phase 3 Success'
;
run;

ods escapechar='^';
options nodate nonumber;
ods graphics / border=no height=3in width=6.0in;


proc sgpanel data=binomial_stack noautolegend;
panelby phase / novarname;
format phase phase.;
refline lower_margin / axis=x lineattrs=(pattern=dot);
series x=diff y=dH_success_ddiff / group=phase lineattrs=(thickness=1);
rowaxis label="Confidence Density" offsetmin=0.02;
colaxis label="True Difference in Proportions" offsetmin=0 offsetmax=0;
footnote1 j=left "^{unicode alpha}=&alpha_phase2. for phase 2 LR test against difference 
<= &lower_margin_phase2. with N=&n_ctrl_phase2. per arm.";
footnote2 j=left "^{unicode alpha}=&alpha_phase3. for phase 3 LR test against difference 
<= &lower_margin_phase3. with N=&n_ctrl_phase3. per arm.";
run;


proc sgpanel data=binomial_stack noautolegend;
panelby phase / novarname;
format phase phase.;
refline lower_margin / axis=x lineattrs=(pattern=dot);
series x=diff y=C_success / group=phase lineattrs=(thickness=1);
rowaxis label="Confidence Curve" offsetmin=0.02;
colaxis label="True Difference in Proportions" offsetmin=0 offsetmax=0;
footnote1 j=left "^{unicode alpha}=&alpha_phase2. for phase 2 LR test against difference 
<= &lower_margin_phase2. with N=&n_ctrl_phase2. per arm.";
footnote2 j=left "^{unicode alpha}=&alpha_phase3. for phase 3 LR test against difference 
<= &lower_margin_phase3. with N=&n_ctrl_phase3. per arm.";
run;



options nodate nonumber;
ods graphics / border=no height=3in width=6.0in;
ods escapechar="^";
proc sgplot data=binomial noautolegend;
refline diff_hat / axis=x lineattrs=(thickness=0.5);
series x=diff y=phase2_power / lineattrs=(thickness=1) name="phase2_power";
series x=diff y=phase3_power / lineattrs=(thickness=2 color=cxD05B5B) 
	name="phase3_power";
series x=diff y=phase2and3_power / lineattrs=(thickness=4 color=cx66A5A0) 
	name="phase2and3_power";
series x=diff y=dHddiff  / lineattrs=(thickness=2 pattern=dash color=black)y2axis 
	name="CD";
keylegend "phase2_power" "phase3_power" "phase2and3_power" "CD";
series x=diff y=ref1 / lineattrs=(color=grey pattern=dot);
series x=diff y=ref2 / lineattrs=(color=grey pattern=dot);
series x=ref3 y=phase2_power / lineattrs=(color=grey pattern=dot);
series x=ref4 y=phase3_power / lineattrs=(color=grey pattern=dot);
y2axis values=(0 to 24 by 4) offsetmin=0.02;
footnote1 j=left "^{unicode alpha}=&alpha_phase2. for phase 2 LR test against difference 
<= &lower_margin_phase2. with N=&n_ctrl_phase2. per arm.";
footnote2 j=left "^{unicode alpha}=&alpha_phase3. for phase 3 LR test against difference 
<= &lower_margin_phase3. with N=&n_ctrl_phase3. per arm.";
xaxis label="True Difference in Proportions" offsetmin=0 offsetmax=0;
yaxis label="Power" offsetmin=0.02;
label phase2_power="Phase 2 Power" phase3_power="Phase 3 Power" 
	phase2and3_power="Phase 2 and 3 Power" dHddiff="Confidence Density";
run;




options nodate nonumber;
ods graphics / border=no height=3in width=6.0in;
ods escapechar="^";
proc sgplot data=binomial noautolegend;
refline diff_hat / axis=x lineattrs=(thickness=0.5);
series x=diff y=phase2_power / lineattrs=(thickness=1) name="phase2_power";
series x=diff y=phase3_power / lineattrs=(thickness=2 color=cxD05B5B) 
	name="phase3_power";
series x=diff y=phase2and3_power / lineattrs=(thickness=4 color=cx66A5A0) 
	name="phase2and3_power";
series x=diff y=C  / lineattrs=(thickness=2 pattern=dash color=black)y2axis 
	name="CD";
keylegend "phase2_power" "phase3_power" "phase2and3_power" "CD";
series x=diff y=ref1 / lineattrs=(color=grey pattern=dot);
series x=diff y=ref2 / lineattrs=(color=grey pattern=dot);
series x=ref3 y=phase2_power / lineattrs=(color=grey pattern=dot);
series x=ref4 y=phase3_power / lineattrs=(color=grey pattern=dot);
y2axis max=1 offsetmin=0.02;
footnote1 j=left "^{unicode alpha}=&alpha_phase2. for phase 2 LR test against difference 
<= &lower_margin_phase2. with N=&n_ctrl_phase2. per arm.";
footnote2 j=left "^{unicode alpha}=&alpha_phase3. for phase 3 LR test against difference 
<= &lower_margin_phase3. with N=&n_ctrl_phase3. per arm.";
xaxis label="True Difference in Proportions" offsetmin=0 offsetmax=0;
yaxis label="Power" offsetmin=0.02;
label phase2_power="Phase 2 Power" phase3_power="Phase 3 Power" 
	phase2and3_power="Phase 2 and 3 Power" C="Confidence Curve";
run;




ods graphics / border=no height=3.5in width=6.0in; 

proc sgplot data=binomial;
series x=phase2_power y=dH_dphase2power / lineattrs=(thickness=1) name="phase2_power";
series x=phase3_power y=dH_dpower / lineattrs=(thickness=2 color=cxD05B5B) 
	name="phase3_power";
series x=phase2and3_power y=dH_dphase23power / lineattrs=(thickness=4 color=cx66A5A0) 
	name="phase23_power";
refline &phase2_power_mle. / axis=x lineattrs=(color=blue pattern=dot) 
	legendlabel="Phase 2 Power mle" name="Phase 2 PoS (Power mle)";
refline &phase3_power_mle. / axis=x lineattrs=(color=cxD05B5B pattern=dot thickness=2) 
	legendlabel="Phase 3 Power mle" name="Phase 3 PoS (Power mle)";
refline &phase23_power_mle. / axis=x lineattrs=(color=cx66A5A0 pattern=dot thickness=4) 
	legendlabel="Phase 2 and 3 Power mle" name="Phase 2 and 3 PoS (Power mle)";
footnote1 j=left "Phase 2 Power: mle=&phase2_power_mle., pos=&mean_phase2_power." ;
footnote2 j=left "Phase 3 Power: mle=&phase3_power_mle., pos=&mean_phase3_power.";
footnote3 j=left "Phase 2 and 3 Power: mle=&phase23_power_mle., 
pos=&mean_phase23_power.";
xaxis label="Power";
yaxis label="Confidence Density" min=0 max=6 offsetmin=0.02;
label dH_dphase2power="Phase 2 CD" dH_dpower="Phase 3 CD" 
	dH_dphase23power="Phase 2 and 3 CD";
keylegend "phase2_power"  "phase3_power"   "phase23_power"  "Phase 2 PoS (Power mle)" 
	"Phase 3 PoS (Power mle)" "Phase 2 and 3 PoS (Power mle)"  ;
run;
footnote;




ods graphics / border=no height=3.5in width=6.0in; 

proc sgplot data=binomial;
series x=phase2_power y=C / lineattrs=(thickness=1) name="phase2_power"
	legendlabel="Phase 2 CC";
series x=phase3_power y=C / lineattrs=(thickness=2 color=cxD05B5B) 
	name="phase3_power" legendlabel="Phase 3 CC";
series x=phase2and3_power y=C / lineattrs=(thickness=4 color=cx66A5A0) 
	name="phase23_power" legendlabel="Phase 2 and 3 CC";
footnote1 j=left "Phase 2 Power: mle=&phase2_power_mle., pos=&mean_phase2_power." ;
footnote2 j=left "Phase 3 Power: mle=&phase3_power_mle., pos=&mean_phase3_power.";
footnote3 j=left "Phase 2 and 3 Power: mle=&phase23_power_mle., 
pos=&mean_phase23_power.";
xaxis label="Power";
yaxis label="Confidence Curve" min=0 max=0.8 offsetmin=0.02;
keylegend "phase2_power"  "phase3_power"   "phase23_power"  "Phase 2 PoS (Power mle)" 
	"Phase 3 PoS (Power mle)" "Phase 2 and 3 PoS (Power mle)"  ;
run;
footnote;






options nodate nonumber;
ods graphics / border=no height=3in width=6.0in;
ods escapechar="^";
proc sgplot data=binomial noautolegend;
band x=diff upper=area lower=0 / fillattrs=(color=lightgrey) y2axis;
refline lower_cv2 / axis=x lineattrs=(thickness=0.5);
series x=diff y=dH_phase2_success_ddiff  / lineattrs=(thickness=1 pattern=solid) 
	y2axis name="CD" legendlabel="Minimum Phase 2 Success";
series x=diff y=phase3_power / lineattrs=(thickness=2 color=cxD05B5B) 
	name="phase3_power";
keylegend "phase2_power" "phase3_power" "phase2and3_power" "CD";
series x=diff y=ref5 / lineattrs=(color=grey pattern=dot);
series x=ref6 y=phase3_power / lineattrs=(color=grey pattern=dot);
y2axis values=(0 to 24 by 4) offsetmin=0.02;
footnote1 j=left "^{unicode alpha}=&alpha_phase2. for phase 2 LR test against difference 
<= &lower_margin_phase2. with N=&n_ctrl_phase2. per arm.";
footnote2 j=left "^{unicode alpha}=&alpha_phase3. for phase 3 LR test against difference 
<= &lower_margin_phase3. with N=&n_ctrl_phase3. per arm.";
xaxis label="True Difference in Proportions" offsetmin=0 offsetmax=0;
yaxis label="Power" offsetmin=0.02;
label phase3_power="Phase 3 Power" 
dH_phase2_success_ddiff="Confidence Density";
run;
footnote;




options nodate nonumber;
ods graphics / border=no height=3in width=6.0in;
ods escapechar="^";
proc sgplot data=binomial noautolegend;
refline lower_cv2 / axis=x lineattrs=(thickness=0.5);
series x=diff y=C_phase2_success  / lineattrs=(thickness=1 pattern=solid) 
	y2axis name="CD" legendlabel="Minimum Phase 2 Success";
series x=diff y=phase3_power / lineattrs=(thickness=2 color=cxD05B5B) 
	name="phase3_power";
keylegend "phase2_power" "phase3_power" "phase2and3_power" "CD";
series x=diff y=ref5 / lineattrs=(color=grey pattern=dot);
series x=ref6 y=phase3_power / lineattrs=(color=grey pattern=dot);
y2axis max=1 offsetmin=0.02;
footnote1 j=left "^{unicode alpha}=&alpha_phase2. for phase 2 LR test against difference 
<= &lower_margin_phase2. with N=&n_ctrl_phase2. per arm.";
footnote2 j=left "^{unicode alpha}=&alpha_phase3. for phase 3 LR test against difference 
<= &lower_margin_phase3. with N=&n_ctrl_phase3. per arm.";
xaxis label="True Difference in Proportions" offsetmin=0 offsetmax=0;
yaxis label="Power" offsetmin=0.02;
label phase3_power="Phase 3 Power" 
C_phase2_success="Confidence Curve";
run;
footnote;






ods graphics / border=no height=3.5in width=6.0in;


proc sgplot data=binomial;
band x=phase3_power upper=area2 lower=0 / fillattrs=(color=lightgrey);
series x=phase3_power y=dH_phase2_dphase3 / lineattrs=(thickness=2 color=cxD05B5B) 
	name="phase3_power";
refline &phase3cond2_power_mle. / axis=x lineattrs=(color=cxD05B5B pattern=dot 
	thickness=2) legendlabel="Phase 3 Power mle" name="Phase 3 Power mle";
footnote1 j=left "Phase 3 Power: mle=%sysfunc(strip(&phase3cond2_power_mle.)), 
pos=&mean_phase3cond2_power.";
xaxis label="Power";
yaxis label="Confidence Density" min=0 max=6 offsetmin=0.02;
label  dH_phase2_dphase3="Phase 3 CD" ;
keylegend "phase3_power"     "Phase 3 Power mle"  ;
run;
footnote;



options nodate nonumber;
ods graphics / border=no height=3.5in width=6.0in; 
proc sgplot data=binomial;
refline lower_cv2 / axis=x lineattrs=(thickness=0.5);
series x=diff y=dH_phase2_success_ddiff / lineattrs=(pattern=solid) name="phase 2" 
	legendlabel="(ii) Minimum Phase 2 Success" y2axis;
series x=diff y=phase3_power / lineattrs=(thickness=2 color=cxD05B5B) 
	name="phase3_power" legendlabel="(v) Phase 3 Power";
series x=diff y=dH_multiply_ddiff / lineattrs=(pattern=8) name="multiply" 
	legendlabel="(iii) Multiplication" y2axis;
series x=diff y=dH_convolve_ddiff / lineattrs=(pattern=3) name="convolution" 
	legendlabel="(iv) Convolution" y2axis;
series x=diff y=dHddiff / lineattrs=(color=black thickness=2 pattern=dash) 
	name="elicited" legendlabel="(i) Elicited Confidence Density" y2axis;
xaxis label="True Difference in Proportions" offsetmin=0 offsetmax=0;
yaxis label="Power" offsetmin=0.02;
y2axis values=(0 to 24 by 4) label="Confidence Density" offsetmin=0.02;
keylegend "elicited" "phase 2" "multiply" "convolution" "phase3_power";
series x=diff y=ref5 / lineattrs=(color=grey pattern=dot);
series x=ref6 y=phase3_power / lineattrs=(color=grey pattern=dot);
footnote1 j=left "^{unicode alpha}=&alpha_phase2. for phase 2 LR test against difference 
<= &lower_margin_phase2. with N=&n_ctrl_phase2. per arm.";
footnote2 j=left "^{unicode alpha}=&alpha_phase3. for phase 3 LR test against difference 
<= &lower_margin_phase3. with N=&n_ctrl_phase3. per arm.";
run;



options nodate nonumber;
ods graphics / border=no height=3.5in width=6.0in; 
proc sgplot data=binomial;
refline lower_cv2 / axis=x lineattrs=(thickness=0.5);
series x=diff y=C_phase2_success / lineattrs=(pattern=solid) name="phase 2" 
	legendlabel="(ii) Minimum Phase 2 Success" y2axis;
series x=diff y=phase3_power / lineattrs=(thickness=2 color=cxD05B5B) 
	name="phase3_power" legendlabel="(v) Phase 3 Power";
series x=diff y=C_multiply / lineattrs=(pattern=8) name="multiply" 
	legendlabel="(iii) Multiplication" y2axis;
series x=diff y=C_convolve / lineattrs=(pattern=3) name="convolution" 
	legendlabel="(iv) Convolution" y2axis;
series x=diff y=C / lineattrs=(color=black thickness=2 pattern=dash) 
	name="elicited" legendlabel="(i) Elicited Confidence Curve" y2axis;
xaxis label="True Difference in Proportions" offsetmin=0 offsetmax=0;
yaxis label="Power" offsetmin=0.02;
y2axis max=1 label="Confidence Curve" offsetmin=0.02;
keylegend "elicited" "phase 2" "multiply" "convolution" "phase3_power";
series x=diff y=ref5 / lineattrs=(color=grey pattern=dot);
series x=ref6 y=phase3_power / lineattrs=(color=grey pattern=dot);
footnote1 j=left "^{unicode alpha}=&alpha_phase2. for phase 2 LR test against difference 
<= &lower_margin_phase2. with N=&n_ctrl_phase2. per arm.";
footnote2 j=left "^{unicode alpha}=&alpha_phase3. for phase 3 LR test against difference 
<= &lower_margin_phase3. with N=&n_ctrl_phase3. per arm.";
run;
 
 
\end{lstlisting}

\end{document}